\documentclass{article}
\usepackage{hyperref}
\usepackage[utf8]{inputenc}
\usepackage[a4paper, total={6.4in, 9.6in}]{geometry}
\usepackage{amssymb}
\usepackage{url}
\usepackage{amsmath}
\usepackage{mathrsfs}
\usepackage{mathtools}
\usepackage{siunitx}
\usepackage{bigints}
\usepackage{longtable}
\usepackage{subfig}
\usepackage{graphicx}
\allowdisplaybreaks

\title{\textbf{Neutron Star properties in $f(\mathcal{R})$ gravity}}
\author{\textbf{Pinaki Roy\footnote{E-mail: pinaki.roy1989@gmail.com}} \\Project Student \\ \\supervised by\\ \textbf{Bharat Kumar}\\ \\Assistant Professor\\National Institute of Technology\\Rourkela, Odisha, India}
\date{ }
\vspace{-3pt}
\begin{document}

\maketitle

\begin{abstract}

In this work, we systematically derive the Einstein field equations in general relativity and $f(\mathcal{R})$ gravity, the Tolman-Oppenheimer-Volkoff (TOV) equation, and the expressions for axial and polar Tidal Love Numbers (TLNs) for neutron stars. The derivations are sourced from existing literature and elaborated for the ease of comprehension.

\end{abstract}

\section*{$f(R)$ Gravity and TOV Equation}
We discuss the Mass–Radius relation for static neutron star models obtained by the numerical solution of (modified) Tolman-Oppenheimer-Volkoff (TOV) equations in $f(R)$ gravity where the Lagrangians adopted are $f(R)=R+\alpha R^2+\beta R^3$ and $f(R)=R^{1+\epsilon}$.\\
 
\noindent
$f(R)$ gravity is a straightforward extension of GR in which the need for the gravity action to be linear in the Ricci scalar, $R$ (scalar curvature), is relaxed as against the Hilbert-Einstein case. It can be seen as a simple case of a more general class of Extended (or Modified) Theories of Gravity. The reason we resort to such an approach is that higher order curvature corrections can emerge in the extreme gravity regimes inside a NS.\\
 
\noindent
The metric for spherically symmetric system has the usual form (we adopt the signature $(-+++)$):
\begin{gather}
\boldsymbol{g_{\mu\nu}=\textbf{diag}(-c^2 e^{2w}, e^{2\lambda}, r^2, r^2 \sin^2{\theta})}\\
\implies ds^2 = -c^2 e^{2w} dt^2 + e^{2\lambda}dr^2 + r^2(d\theta^2 + \sin^2{\theta}\,d\phi^2)\nonumber\\
\Leftrightarrow ds^2 = -c^2 e^{2w} dt^2 + e^{2\lambda}dr^2 + r^2d\Omega^2\nonumber
\end{gather}

\noindent
where $w$ and $\lambda$ are \textbf{dimensionless} functions depending only on the radial coordinate $r$. The stellar matter is described as a perfect fluid, whose energy-momentum tensor (or, stress-energy tensor) is $\boldsymbol{T_{\mu\nu}=\textbf{diag}(e^{2w}\rho c^4, e^{2\lambda}p, r^2 p, r^2 p \sin^2{\theta})}$, where $\rho$ is the \textbf{matter density} and $p$ is the pressure.\\

\noindent
The stellar structure equations are obtained on addition of the energy-momentum tensor continuity (or, conservation) equation.
\begin{equation}
\nabla_{\mu}T^{\mu\nu}=0\implies\boldsymbol{\dfrac{dp}{dr}=-(\rho c^2+p)\dfrac{dw}{dr}}
\end{equation}

\noindent
The $f(R)$ theories of gravity are defined by the action:
$$\mathcal{S}=\dfrac{1}{2\kappa}\displaystyle\int d^4x\sqrt{-g}\,f(R)+\mathcal{S}_{\text{M}}$$
$$\Leftrightarrow\mathcal{S}=\displaystyle\int d^4x\sqrt{-g}\,\left[\dfrac{1}{2\kappa}f(R)+\mathcal{L}_{\text{M}}\right]$$

\noindent
where $\mathcal{S}_{\text{M}}$ and $\mathcal{L}_{\text{M}}$ are the matter action and the matter Lagrangian respectively, $\boldsymbol{\kappa=8\pi G/c^4}$, and $g$ is the determinant (necessarily negative) of the metric $g_{\mu\nu}$ i.e. $\boldsymbol{g=\textbf{det}\,(g_{\mu\nu})}$. For the static metric in (1), $\boldsymbol{g=-c^2e^{(2w+2\lambda)}r^4\sin^2{\theta}}$ so that $\boldsymbol{\sqrt{-g}=ce^{(w+\lambda)}r^2\sin{\theta}}$\\

\noindent
The variation of the action with respect to $g_{\mu\nu}$ gives the (modified) field equations:
\begin{gather}
\boldsymbol{f_RR_{\mu\nu}-\dfrac{1}{2}f(R)g_{\mu\nu}-\left(\nabla_{\mu}\nabla_{\nu}-g_{\mu\nu}\Box\right)f_R=\dfrac{8\pi G}{c^4}T_{\mu\nu}}\\
\Leftrightarrow
f_R\,\mathcal{G}_{\mu\nu}- \dfrac{1}{2}\left(f(R)-Rf_R\right)g_{\mu\nu}+\left(\nabla_{\mu}\nabla_{\nu}-g_{\mu\nu}g^{\alpha\beta}\nabla_{\alpha}\nabla_{\beta}\right)f_R=\dfrac{8\pi G}{c^4}T_{\mu\nu}\\
\Leftrightarrow \boldsymbol{\left(R_{\mu\nu}-\nabla_{\mu}\nabla_{\nu}\right)f_R-\dfrac{1}{2}g_{\mu\nu}\left(f(R)-2\,\Box f_R\right)=\dfrac{8\pi G}{c^4}T_{\mu\nu}}\nonumber
\end{gather}

\noindent
where $\Box=g^{\alpha\beta}\nabla_{\alpha}\nabla_{\beta}$ is the d'Alembertian (operator) in curved spacetime, $T_{\mu\nu}:=\dfrac{-2}{\sqrt{-g}}\dfrac{\delta\left(\sqrt{-g}\,\mathcal{L}_\text{M}\right)}{\delta g^{\mu\nu}}=-2\dfrac{\delta\mathcal{L}_\text{M}}{\delta g^{\mu\nu}}+g_{\mu\nu}{L}_\text{M}$, Einstein tensor, $\boldsymbol{\mathcal{G}_{\mu\nu}=R_{\mu\nu}-\dfrac{1}{2}Rg_{\mu\nu}}$ and $\boldsymbol{f_R=\dfrac{df(R)}{dR}}$ $\left(\text{similarly, } f_{RR}=\dfrac{df_R}{dR}\right)$\\\\

\noindent
The covariant derivative of a scalar $\psi$ is a covariant vector given by $$\nabla_{\beta}\psi=\dfrac{\partial}{\partial x^{\beta}}\psi\equiv\partial_{\beta}\psi$$

\noindent
The covariant derivative of a covariant vector $V_{\mu}$ is a covariant rank-2 tensor given by $$\nabla_{\alpha}V{_\beta}=\dfrac{\partial}{\partial x^{\alpha}}V_{\beta}-\Gamma_{\alpha\beta}^{\lambda}V{_\lambda}\equiv\partial_\alpha V_{\beta}-\Gamma_{\alpha\beta}^{\lambda}V{_\lambda}$$

\noindent
so that $\nabla_{\alpha}\nabla_{\beta}\psi=\dfrac{\partial}{\partial x^{\alpha}}\dfrac{\partial}{\partial x^{\beta}}\psi-\Gamma_{\alpha\beta}^{\lambda}\dfrac{\partial}{\partial x^{\lambda}}\psi \equiv \left(\partial_{\alpha}\partial_{\beta}-\Gamma_{\alpha\beta}^{\lambda}\partial_{\lambda}\right)\psi$\\

\noindent
Therefore:
\begin{gather}
\Box f_R= g^{\alpha\beta}\nabla_{\alpha}\nabla_{\beta}f_R=g^{\alpha\beta}\left(\partial_{\alpha}\partial_{\beta}-\Gamma_{\alpha\beta}^{\lambda}\partial_{\lambda}\right)f_R\nonumber\\
\implies \boldsymbol{\Box f_R=e^{-2\lambda}\left[\left(w'-\lambda'+\dfrac{2}{r}\right)\dfrac{df_R}{dr}+\dfrac{d^2f_R}{dr}\right]}
\end{gather}

\noindent
\textbf{Note:} In the above four equations (before (5)) and (4), $\alpha$, $\beta$ and $\lambda$ are dummy indices not to be confused with $\alpha$, $\beta$ and $\lambda$ used elsewhere.
\begin{gather}
\left(\nabla_{\mu}\nabla_{\nu}-g_{\mu\nu}\Box\right)f_R=\nabla_{\mu}\nabla_{\nu}f_R-g_{\mu\nu}\Box f_R
\end{gather}

\noindent
The 00 (i.e. $tt$) component of (6) becomes:
\begin{gather*}
\left(\nabla_{0}\nabla_{0}-g_{00}\Box\right)f_R=-c^2 e^{2w-2\lambda}w'\dfrac{df_R}{dr}+c^2 e^{2w-2\lambda}\left[\left(w'-\lambda'+\dfrac{2}{r}\right)\dfrac{df_R}{dr}+\dfrac{d^2f_R}{dr}\right]\\
\implies \nabla_{0}\nabla_{0}f_R-g_{00}\Box f_R=c^2 e^{2w-2\lambda}\left[\left(-\lambda'+\dfrac{2}{r}\right)\dfrac{df_R}{dr}+\dfrac{d^2f_R}{dr}\right]
\end{gather*}

\noindent
The 11 (i.e. $rr$) component of (6) becomes:
\begin{gather*}
\left(\nabla_{1}\nabla_{1}-g_{11}\Box\right)f_R=\left[\dfrac{d^2 f_R}{dr^2}-\lambda'\dfrac{df_R}{dr}\right]-\left[\left(w'-\lambda'+\dfrac{2}{r}\right)\dfrac{df_R}{dr}+\dfrac{d^2f_R}{dr}\right]\nonumber\\
\implies \nabla_{1}\nabla_{1}f_R-g_{11}\Box f_R=\left(w'+\dfrac{2}{r}\right)\dfrac{df_R}{dr}
\end{gather*}

\noindent
The 00 (i.e. $tt$) component of (4) turns out to be:
\begin{gather}
\dfrac{f_R}{r^2}\dfrac{d}{dr}\left[r\left(1-e^{-2\lambda}\right)\right]-\dfrac{1}{2}\left(f(R)-Rf_R\right)-e^{-2\lambda}\left[\left(\dfrac{2}{r}-\dfrac{d\lambda}{dr}\right)\dfrac{df_R}{dr}+\dfrac{d^2f_R}{dr^2}\right]=\dfrac{8\pi G}{c^4}\rho c^2
\end{gather}

\noindent
The 11 (i.e. $rr$) component of (4) turns out to be:
\begin{gather}
\dfrac{f_R}{r}\left[2e^{-2\lambda}\dfrac{dw}{dr}-\dfrac{1}{r}\left(1-e^{-2\lambda}\right)\right]-\dfrac{1}{2}\left(f(R)-Rf_R\right)-e^{-2\lambda}\left(\dfrac{2}{r}+\dfrac{dw}{dr}\right)\dfrac{df_R}{dr}=\dfrac{8\pi G}{c^4}p
\end{gather}

\noindent
Note that the temporal metric potential $w$ appears $\left(\text{as } \dfrac{dw}{dr}\right)$ in the $rr$ component (8), whereas, the spatial (radial) metric potential $\lambda$ appears $\left(\text{as } \dfrac{d\lambda}{dr}\right)$ in the $tt$ component (7). Equations (7) \& (8) are the \textbf{TOV equations for} $\boldsymbol{f(R)}$ \textbf{gravity}. They reduce to the ordinary TOV equations of GR for $f(R)=R$ in which case $f_R=\dfrac{df(R)}{dR}=1$ so that they respectively become:
\begin{gather*}
\dfrac{1}{r^2}\dfrac{d}{dr}\left[r\left(1-e^{-2\lambda}\right)\right]=\dfrac{8\pi G}{c^4}\rho c^2\\
\dfrac{1}{r}\left[2e^{-2\lambda}\dfrac{dw}{dr}-\dfrac{1}{r}\left(1-e^{-2\lambda}\right)\right]=\dfrac{8\pi G}{c^4}p
\end{gather*}

\noindent
Ricci Scalar (or, Scalar Curvature) for the metric in (1) is
\begin{gather}
R=2e^{-2\lambda}\left[\dfrac{1}{r^2}(e^{2\lambda}-1)-\dfrac{2}{r}(w'-\lambda')-w"-w'^2+w'\lambda'\right]
\end{gather}

\noindent
Taking the trace of the field equations (3), we get the equation for $R$ which is a dynamic variable in $f(R)$ gravity:
\begin{equation}
\boldsymbol{3\,\Box f_R+f_R R-2f(R)=-\dfrac{8\pi G}{c^4}(\rho c^2-3p)}
\end{equation}

\noindent
In GR, $f(R)=R$ so that $f_R=1$ and $\Box f_R=0$, and the equation (4) becomes the trace of GR:
$$R=\dfrac{8\pi G}{c^4}(\rho c^2-3p)$$

\noindent
The unknowns are $w$, $\lambda$, $p$, $\rho$ and $R$. Equations (2) and (9) are two of the five equations required to complete the system. An equation of state (EoS) i.e. $p=p(\rho)$ relating the pressure and the density of the stellar interior would be one more equation. Remaining two equations are (7) and (6). Alternatively, (7) and (6) can be respectively written in the form:
\begin{gather}
\dfrac{dw}{dr}\left[\dfrac{2}{r}f_R-\dfrac{df_R}{dr}\right]e^{-2\lambda}=\dfrac{8\pi G}{c^4}p+\dfrac{1}{2}\left(f(R)-Rf_R\right)+\dfrac{f_R}{r^2}\left(1-e^{-2\lambda}\right)+\dfrac{2}{r}\dfrac{df_R}{dr}e^{-2\lambda}\\
\dfrac{d\lambda}{dr}\left[\dfrac{2}{r}f_R+\dfrac{df_R}{dr}\right]e^{-2\lambda}=\dfrac{8\pi G}{c^4}\rho c^2 +\dfrac{1}{2}\left(f(R)-Rf_R\right)-\dfrac{f_R}{r^2}\left(1-e^{-2\lambda}\right)+\left[\dfrac{2}{r}\dfrac{df_R}{dr}+\dfrac{d^2f_R}{dr^2}\right]e^{-2\lambda}
\end{gather}
Observe that $R\equiv R(w,w',w'',\lambda,\lambda')$ and also $f_R=\dfrac{df(R)}{dR}\equiv f_R(w,w',w'',\lambda,\lambda')$. Equations (2), (9), (10), (11) and EoS form a complete system to determine the five unknowns as they respectively relate $(w,p,\rho)$, $(p,\rho,R)$, $(w,p,R)$, $(\lambda,\rho,R)$ and $(p,\rho)$.\\\\
\textbf{Note:} Metric potentials $w$ and $\lambda$, Pressure and Density, Scalar curvature and Mass coordinate (each of these six) are to be plotted against Radius.\\\\
Before going through the various cases of the theory at hand, it should satisfy two inequalities in order to be free of tachyonic instabilities and ghosts: $f_R\geq 0\text{ and }f_{RR}\geq 0$\\
\textbf{Case I:} $f(R)=R+\alpha R^2$\\
In this case, $f_R=1+2\alpha R$ where $\alpha$ is a free parameter. By (13), $\alpha\geq 0$. \\\\
\textbf{Case II:} $f(R)=R+\alpha R^2+\beta R^3$\\
We also consider a cubic form of $f(R)$ with unspecified coefficients. In this case, $f_R=1+2\alpha R+3\beta R^2$ where $\alpha$ and $\beta$ are a free parameter.\\\\
\textbf{Case III:} $f(R)=R+\beta R^3$\\
This is to explore the possibility of a cubic form of $f(R)$ with no quadratic term. In this case, $f_R=1+3\beta R^2$ where $\beta$ is a free parameter.\\\\
\textbf{Case IV:} $f(R)=R^{1+\epsilon}$\\
Assuming a small deviation with respect to GR, i.e. $|\epsilon| \ll 1$, we have the case: $f(R)=R^{1+\epsilon}=e^{(1+\epsilon)\ln{R}}\simeq R+\epsilon R\ln{R}+\mathcal{O}(\epsilon^2)$ to first-order Taylor expansion (at $\epsilon=0$). In this case, $f_R=1+\epsilon+\epsilon \ln{R}$ where $\epsilon$ is a free parameter.

\subsection*{Neutron Star Structure and EoS}
A polytropic EoS is defined by $p=K\rho^{1+1/n}=K\rho^{\Gamma}$ where $\Gamma=1+\dfrac{1}{n}$ is the polytropic exponent, conventionally called the adiabatic index, and $n$ is referred as the polytropic index. Polytropes are special cases of barotropes, the EoS for which is $\rho=\rho(p)$ or equivalenty, $p=p(\rho)$. So, polytropes are power-law barotropes.\\\\
In this work, we use a unified soft EoS, namely SLy4. Beyond the fixed crust (solid outer crust and inner crust), it can be described as a three-segment piecewise polytrope with four parameters, to wit, $p_1$, $\Gamma_1$, $\Gamma_2$ and $\Gamma_3$ where subscripts (of the adiabatic index) 1, 2 and 3 represent regions 1 (mantle), 2 (outer core) and 3 (inner core) of the star. $\rho_1=10^{14.7}$ g/cm$^3$ and $\rho_2=10^{15.0}$ g/cm$^3$ are the two dividing densities of the three segments. $p_1$ is the pressure at first dividing density (i.e. corresponding to $\rho_1$). Instead of $p_1$, $\log{p_1}$ can also be used as the parameter. The outer crust is taken to begin with the density, $\rho=10^{8.22}\approx1.66\times10^8$ g/cm$^3$ and the inner crust is taken to begin with the density, $\rho=10^{11.6}\approx4\times10^{11}$ g/cm$^3$. Above the crust are the layers of ocean/envelope ($\rho\leq 10^8$ g/cm$^3$) and atmosphere ($\rho\leq 10^4$ g/cm$^3$). The EoS table gives the pressure corresponding to different values of baryon number density and baryon mass density (number density times atomic mass unit) inside the NS.\\\\
Saturation density, $\rho_0=0.74\rho_\text{nuc}=0.04\rho_\text{c}$\\
Region 1 ($10^{14.3}\approx0.74\rho_\text{nuc}\approx2\times10^{14}$ g/cm$^3$ -- $5\times10^{14}$ g/cm$^3\approx1.85\rho_\text{nuc}\approx 10^{14.7}$): $p=K_1\rho^{\Gamma_1}$\\
Region 2 ($10^{14.7}\approx1.85\rho_\text{nuc}\approx5\times10^{14}$ g/cm$^3$ -- $1\times10^{15}$ g/cm$^3\approx3.70\rho_\text{nuc}\approx10^{15.0}$): $p=K_2\rho^{\Gamma_2}$\\
Region 3 ($10^{15.0}\approx3.70\rho_\text{nuc}\approx1\times10^{15}$ g/cm$^3$ -- $5\times10^{15}$ g/cm$^3\approx18.5\rho_\text{nuc}\approx10^{15.7}$): $p=K_3\rho^{\Gamma_3}$\\
Central density, $\rho_\text{c}=18.5\rho_\text{nuc}=25\rho_0$ (this limit is the limit of the EoS in use)\\\\
At boundary 1 (between regions 1 and 2): $K_1\rho_1^{\Gamma_1}=K_2\rho_1^{\Gamma_2}\implies K_2=K_1 \rho_1^{\Gamma_1-\Gamma_2}$\\
At boundary 2 (between regions 2 and 3): $K_2\rho_2^{\Gamma_2}=K_3\rho_2^{\Gamma_3}\implies K_3=K_2 \rho_2^{\Gamma_2-\Gamma_3}$\\\\
Now, $\log{K_1}=\log{p_1}-\Gamma_1 \log{\rho_1}$\\
$\log{K_2}=\log{K_1}+(\Gamma_1-\Gamma_2)\log{\rho_1}=(\log{p_1}-\Gamma_1 \log{\rho_1})+(\Gamma_1-\Gamma_2)\log{\rho_1}=\log{K_1}=\log{p_1}-\Gamma_2 \log{\rho_1}$\\
$\log{K_3}=\log{K_2}+(\Gamma_2-\Gamma_3)\log{\rho_2}=(\log{p_1}-\Gamma_2 \log{\rho_1})+(\Gamma_2-\Gamma_3)\log{\rho_2}\\
\hspace*{35ex}=\log{p_1}+(\log{\rho_2}-\log{\rho_1})\Gamma_2-\Gamma_3\log{\rho_2}$\\\\
To determine the four parameters, $\log{p_1}$, $\Gamma_1$, $\Gamma_2$ and $\Gamma_3$, of the piecewise polytrope model, we minimize the residual (or, its square) over the $N$ tabulated data points, where $N=n(j_1)+n(j_2)+n(j_3)$
$$\text{res}=\sqrt{\dfrac{1}{N}\sum_{i}\sum_{j}(\log{p_j}-\log{K_i}+\Gamma_i\log{\rho_j})^2}$$
\vspace{-2.0em}
\begin{figure}[h!]
\centering
\includegraphics[width=0.95\textwidth]{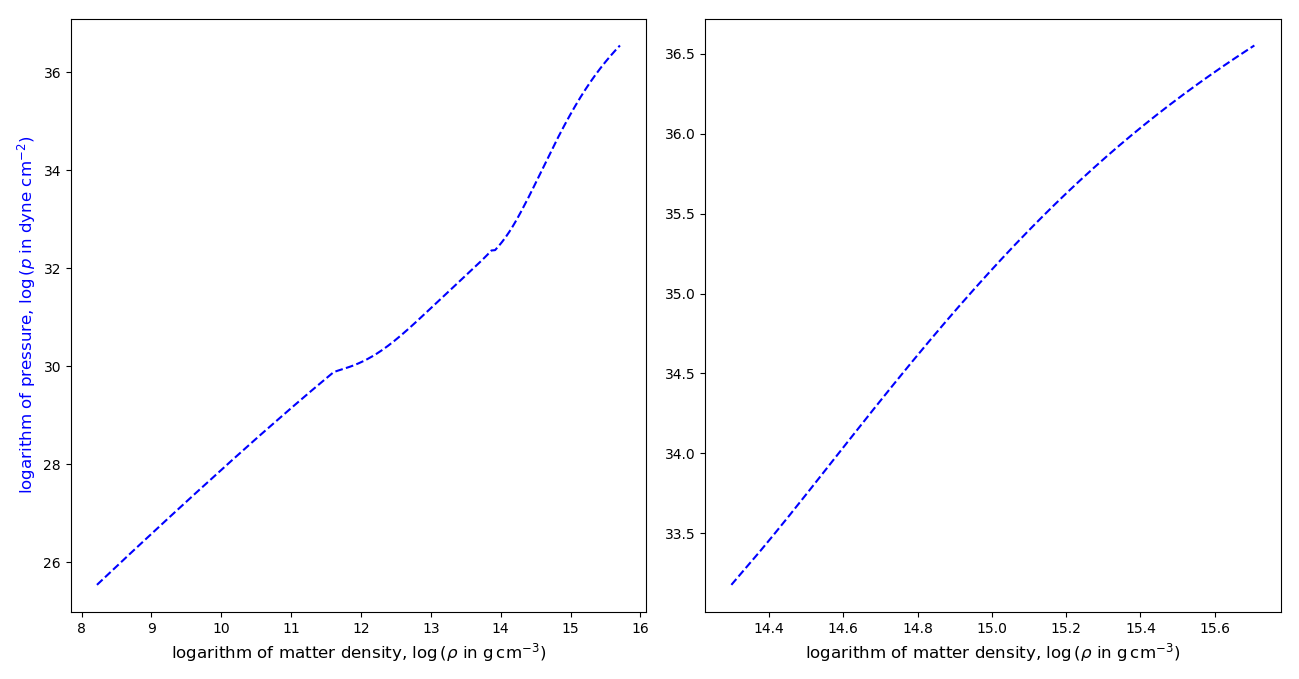}
\caption{(a) SLy4 EoS from crust to center (b) SLy4 EoS beneath the crust}
\end{figure}
\subsection*{Tidal Deformability and Tidal Love Numbers (TLNs)}
Consider the stationary perturbation of the static and spherically symmetric metric in $f(R)$ gravity due to a static external quadrupolar tidal field. The perturbed metric is then:
\begin{equation}
\Tilde{g}_{\mu\nu}=g_{\mu\nu}^{(0)}+h_{\mu\nu} 
\end{equation}
where $h_{\mu\nu}$ is a linearized metric perturbation, and $g_{\mu\nu}^{(0)}=g_{\mu\nu}$ is the unperturbed metric.\\
For polar perturbations, the metric perturbation in the Regge-Wheeler gauge takes the form:
\begin{equation}
h_{\mu\nu}^\text{polar}=\begin{pmatrix}-c^2 e^{2w}H_0(r) & H_1(r) & 0 & 0\\ H_1(r) & e^{2\lambda}H_2(r) & 0 & 0\\0 & 0 & r^2K(r) & 0\\0 & 0 & 0 & r^2K(r) \sin^2{\theta} \end{pmatrix}Y_{lm}(\theta,\phi)
\end{equation}
where $Y_{lm}$ is the spherical harmonic function of degree $l$ and order $m$. It can be shown that $H_1(r)=0$, so that $h_{\mu\nu}^\text{polar}=\text{diag}[-c^2 e^{2w}H_0(r), e^{2\lambda}H_2(r), r^2K(r), r^2K(r) \sin^2{\theta}]\,Y_{lm}(\theta,\phi)$. The tidal deformability, $\eta$, is the factor which relates the star’s tidal induced multipole moment, $\mathcal{I}$, to the external tidal field, $\mathcal{E}$.
\begin{equation}
\mathcal{I}_{ij}=-\eta \mathcal{E}_{ij}
\end{equation}
where $\mathcal{I}_{ij}$ and $\mathcal{E}_{ij}$ are both ($3 \times 3$) symmetric and traceless (STF: symmetric tracefree) tensors, and are defined to be coefficients in an asymptotic expansion of the total metric at large distances, $r$, from the star.
\begin{gather}
\Tilde{g}_{tt}=-c^2\bigg(1-\dfrac{2GM}{c^2r}\bigg)+\dfrac{3G\mathcal{I}_{ij}}{r^3}n^i n^j-\mathcal{E}_{ij} n^i n^jr^2+\dots
\end{gather}
with $n^i=x^i/r=(\sin{\theta}\cos{\phi},\sin{\theta}\sin{\phi},\cos{\theta})$ is the unit radial vector, and $\mathcal{E}_{ij}=\mathcal{R}_{i0j0}$. Now, $\mathcal{I}_{ij}$ and $\mathcal{E}_{ij}$ can be decomposed as\,:
\begin{gather}
\mathcal{I}_{ij}=\displaystyle\sum_{m=-l}^{l}\mathcal{I}_{m}\mathcal{Y}_{ij}^{lm}\quad \text{and} \quad \mathcal{E}_{ij}=\displaystyle\sum_{m=-l}^{l}\mathcal{E}_{m}\mathcal{Y}_{ij}^{lm}
\end{gather}
where the STF tensors, $\mathcal{Y}_{ij}^{lm}$ are defined as
\begin{gather}
Y_{lm}(\theta,\phi)=\mathcal{Y}_{ij}^{lm}n^in^j
\end{gather}
Thus, equation (15) becomes\,:
\begin{equation}
\mathcal{I}_{m}=-\eta_l \mathcal{E}_{m}
\end{equation}
The (dimensionless) second tidal Love number for degree $l$ is
\begin{equation}
k_l=\dfrac{(2l-1)!!G\eta_l}{2R^{2l+1}}
\end{equation}
so that for $l=$2 (quadrupole), 3 (octapole) and 4 (hexadecupole),
\begin{equation}
k_2=\dfrac{3G\eta_2}{2R^5}\qquad\qquad k_3=\dfrac{15G\eta_3}{2R^7}\qquad\qquad k_4=\dfrac{105G\eta_4}{2R^9}
\end{equation}
\subsubsection*{Calculation of polar TLN}
\textbf{Objective}: to obtain expression for the $l=2$ polar tidal Love number, $k_2(H,H')$, and the solution for $H$\\
\textbf{Ingredients:} the form of the metric, perturbation of the metric (in Regge-Wheeler gauge), TOV equations, continuity equation of stress-energy tensor, asymptotic expansion of the metric\\
\textbf{Prescription:} It is convenient to substitute $\cos^{-1}{z}$ for $\theta$ while defining the metric and the perturbation at the beginning. 1. From the perturbation of the metric and the stress-energy tensor, obtain the linearized EFE, $\delta G_{\mu}^{\,\,\nu}=\kappa\,\delta T_{\mu}^{\,\,\nu}$. 2. From $\delta G_{\theta}^{\,\,\theta}-\delta G_{\phi}^{\,\,\phi}=0$, deduce $H_0=-H_2\equiv H$. 3. From $\delta G_{\theta}^{\,\,r}=0$, find $K'$ in terms of $H$ and $H'$. 4. Assuming barotropic EoS, express $\delta \rho$ in terms of $\delta p$, and use $\delta G_{\theta}^{\,\,\theta}+\delta G_{\phi}^{\,\,\phi}=2\kappa\,\delta p$ to eliminate $\delta p$. 5. From $\delta G_{t}^{\,\,t}-\delta G_{r}^{\,\,r}=-\kappa(\psi+1)$, obtain the (second order) differential equation for $H$ ($H$-equation). 6. Obtain the exterior form of the $H$-equation and get the analytical solution having transformed into the form of the associated Legendre equation. 7. From the asymptotic behaviour of the exterior solution, determine the coefficients of the exterior solution. 8. Solve for $k_2$ in terms of $H'/H$ at the star's surface $r=R$. 9. For the interior case, split the $H$-equation into two first order differential equations. 10. Solve the coupled differential equations numerically (along with the appropriate TOV equations). 11. Match the interior solution with the exterior solution as well as their first derivatives at $r=R$.
$$R_{\mu}^{\,\,\nu}-\dfrac{1}{2}R\delta_{\mu}^{\,\,\nu}=\dfrac{8\pi G}{c^4}T_{\mu}^{\,\,\nu}\implies G_{\mu}^{\,\,\nu}=\dfrac{8\pi G}{c^4}T_{\mu}^{\,\,\nu}$$
\vspace{-1.5em}
\begin{flalign*}
&T^{\,\,\nu}_{\mu}=\left(\rho+\dfrac{p}{c^2}\right)u_{\mu}u^{\nu}+p\,\delta^{\,\,\nu}_{\mu}\implies \delta T_{\mu}^{\,\,\nu}=-\left(\delta\rho+\dfrac{\delta p}{c^2}\right)c^2+\delta p\,\delta^{\,\,\nu}_{\mu}\qquad[\because \delta(u_{\mu}u^{\nu})=0 \text{ and } \delta\delta^{\,\,\nu}_{\mu}=0]&\\
&\therefore\,\delta T_{\mu}^{\,\,\nu}=\text{diag}(-c^2\delta \rho,\,\delta p,\,\delta p,\,\delta p)=\text{diag}(-\psi(r),\,1,\,1,\,1)\delta p\qquad[\,\text{where }\psi(r)=c^2 d\rho/dp,]&
\end{flalign*}
The expression for $K'$ turns out to be (prime denotes differentiation w.r.t. $r$):
$$K'=-H'-2w'H\implies K=-H-2\left(wH-\displaystyle\int{wH'dr}\right)=-(1+2w)H+\displaystyle\int{wH'dr}$$
The differential equation for $H(r)$ for $l=2$ and $m=0$ is:
\begin{gather}
H''+C_0 H'+C_1 H=0
\end{gather}
which is a second order linear homogeneous ordinary differential equation with non-constant coefficients.\\\\
where $C_0=\bigg(\dfrac{2}{r}+w'-\lambda'\bigg)=\bigg[\dfrac{2}{r}+\dfrac{G}{c^2}e^{2\lambda}\bigg(\dfrac{2m}{r^2}+4\pi r\Big(\dfrac{p}{c^2}-\rho\Big)\bigg)\bigg]$\\
\hspace*{4em}$=e^{2\lambda}\bigg[\dfrac{2}{r}\bigg(1-\dfrac{2Gm}{c^2 r}\bigg)+\dfrac{G}{c^2}\bigg(\dfrac{2m}{r^2}+4\pi r\Big(\dfrac{p}{c^2}-\rho\Big)\bigg)\bigg]$\\
\hspace*{4em}$=e^{2\lambda}\bigg[\dfrac{2}{r}-\dfrac{G}{c^2}\bigg(\dfrac{2m}{r^2}+4\pi r\Big(\rho-\dfrac{p}{c^2}\Big)\bigg)\bigg]\qquad[\text{using eqns. } (25), (26) \text{ and } (27)]$\\
and $C_1=-\dfrac{6e^{2\lambda}}{r^2}+\dfrac{1}{r}(7w'+3\lambda')+\dfrac{\psi}{r}(w'+\lambda')-2w'(w'+\lambda')+2w''$\\
\hspace*{3.5em}$=-\dfrac{6e^{2\lambda}}{r^2}+\dfrac{3}{r}(w'+\lambda')+\dfrac{4\pi G}{c^2}e^{2\lambda}\psi\bigg(\dfrac{p}{c^2}+\rho\bigg)-4w'^2+2w'(w'+\lambda')+\dfrac{8\pi G}{c^2}e^{2\lambda}\bigg(\rho+\dfrac{3p}{c^2}-rw'(\rho+\dfrac{p}{c^2})\bigg)$\\
\hspace*{3.5em}$=e^{2\lambda}\bigg[-\dfrac{6}{r^2}+\dfrac{12\pi G}{c^2}\bigg(\dfrac{p}{c^2}+\rho\bigg)+\dfrac{4\pi G}{c^2}\psi\bigg(\dfrac{p}{c^2}+\rho\bigg)+\dfrac{8\pi G}{c^2}\bigg(\rho+\dfrac{3p}{c^2}\bigg)\bigg]-4w'^2$\\
\hspace*{3.5em}$=e^{2\lambda}\bigg[-\dfrac{6}{r^2}+4\pi G\bigg(\rho+\dfrac{p}{c^2}\bigg)\dfrac{d\rho}{dp}+\dfrac{4\pi G}{c^2}\bigg(5\rho+\dfrac{9p}{c^2}\bigg)\bigg]-4w'^2\quad[\text{using eqns. } (23), (24), (26), (27) \text{ and } (28)]$\\\\
The TOV and associated equations are:
\begin{gather}
m'=4\pi r^2 \rho\\
w'=-\dfrac{p'}{\rho c^2+p}\\
e^{2\lambda}=\bigg(1-\dfrac{2Gm}{c^2 r}\bigg)^{-1}\\
\implies\lambda'=\dfrac{G}{c^2}e^{2\lambda}\bigg(-\dfrac{m}{r^2}+\dfrac{m'}{r}\bigg)=\dfrac{G}{c^2r^2}e^{2\lambda}\bigg(-m+4\pi r^3\rho\bigg)\\
w'=\dfrac{G}{c^2r^2}e^{2\lambda}\bigg(m+\dfrac{4\pi r^3}{c^2}p\bigg)\\
\implies w''=-\dfrac{2w'}{r}+2\lambda'w'+\dfrac{4\pi G}{c^2}e^{2\lambda}\bigg(\rho+\dfrac{3p}{c^2}-rw'(\rho+\dfrac{p}{c^2})\bigg)\\
p'=-\dfrac{G}{c^2 r^2}e^{2\lambda}(\rho c^2 +p)\bigg(m+\dfrac{4\pi r^3}{c^2}p\bigg)\\
\implies p''=-\dfrac{G}{c^2 r^2}e^{2\lambda}(\rho c^2 +p)\bigg(m+\dfrac{4\pi r^3}{c^2}p\bigg)\left(-\dfrac{2}{r}+2\lambda'+\dfrac{\rho' c^2 +p'}{\rho c^2 +p}+\dfrac{m'+\dfrac{4\pi r^2}{c^2}(rp'+3p)}{m+\dfrac{4\pi r^3}{c^2}p}\right)
\end{gather}
Outside the star, i.e. at $r\geq R$, $p=\rho=0$ and $m=M$, so that $w'=-\lambda'$ which implies $C_0=\dfrac{2}{r}-2\lambda'$ and, $C_1=-\dfrac{6}{r^2}e^{2\lambda}-4\lambda'^2$ where $e^{2\lambda}=\bigg(1-\dfrac{2GM}{c^2 r}\bigg)^{-1}$ and $\lambda'=-\dfrac{GM}{c^2 r}e^{2\lambda}$. The $H$-equation then becomes
\begin{gather}
H''+\bigg(\dfrac{2}{r}-2\lambda'\bigg)H'-\bigg(\dfrac{6}{r^2}e^{2\lambda}+4\lambda'^2\bigg)H=0\\
\implies H''+\dfrac{2}{r}\bigg(1+\dfrac{GM}{c^2r}e^{2\lambda}\bigg)H'-\dfrac{e^{2\lambda}}{r^2}\bigg(6+4\bigg(\dfrac{GM}{c^2r}\bigg)^2 e^{2\lambda}\bigg)H=0\nonumber\\
\implies \bigg(1-\dfrac{2GM}{c^2 r}\bigg)H''+\dfrac{2}{r}\bigg(1-\dfrac{GM}{c^2 r}\bigg)H'-\dfrac{1}{r^2}\bigg(6+\dfrac{4\bigg(\dfrac{GM}{c^2 r}\bigg)^2}{1-\dfrac{2GM}{c^2 r}}\bigg)H=0\nonumber\\
\implies \bigg(\dfrac{c^2 r}{GM}-2\bigg)H''+\dfrac{2}{r}\bigg(\dfrac{c^2 r}{GM}-1\bigg)H'-\dfrac{1}{r^2}\bigg\{\dfrac{6c^2r}{GM}+4\bigg(\dfrac{c^2 r}{GM}-2\bigg)^{-1}\bigg\}H=0\nonumber
\end{gather}
Now, put $x=\bigg(\dfrac{c^2 r}{GM}-1\bigg)$ i.e. $r=\dfrac{GM}{c^2}(x+1)$ so that $\dfrac{d}{dr}=\dfrac{dx}{dr}\dfrac{d}{dx}=\bigg(\dfrac{c^2}{GM}\bigg)\dfrac{d}{dx}$ and $\dfrac{d^2}{dr^2}=\bigg(\dfrac{c^2}{GM}\bigg)^2\dfrac{d^2}{dx^2}$\\
\begin{gather}
\therefore\,(x-1)H''+\dfrac{2x}{x+1}H'-\dfrac{1}{(x+1)^2}\bigg(6(x+1)+\dfrac{4}{x-1}\bigg)H=0\quad[\,\text{where prime denotes differentiation w.r.t. }x\,]\nonumber\\
\implies (x^2-1)H''+2xH'-\bigg(6+\dfrac{4}{x^2-1}\bigg)H=0
\end{gather}
which is an associated Legendre equation with $l'=l=2$ and $m'=2$ whose solution is
\begin{gather}
H(x)=c_P P_2^2(x)+c_Q Q_2^2(x)
\end{gather}
where $P_{l'}^{m'}(x)$ and $Q_{l'}^{m'}(x)$ are associated Legendre functions of first kind and second kind respectively.
\begin{flalign*}
&P_2^2(x)=-3(x^2-1)=-3(x+1)(x-1)\implies P_2^2(r)=-3\bigg(\dfrac{c^2 r}{GM}\bigg)\bigg(\dfrac{c^2 r}{GM}-2\bigg)=-3\bigg(\dfrac{c^2 r}{GM}\bigg)^2\bigg(1-\dfrac{2GM}{c^2 r}\bigg)&\\
&Q_2^2(x)=\dfrac{x(3x^2-5)}{x^2-1}+3(x^2-1)\ln{\sqrt{\dfrac{x-1}{x+1}}}=(x^2-1)\left[\dfrac{x(3x^2-5)}{(x^2-1)^2}+\dfrac{3}{2}\ln{\dfrac{x-1}{x+1}}\right]\\
&\implies Q_2^2(r)=\bigg(\dfrac{c^2 r}{GM}\bigg)^2\bigg(1-\dfrac{2GM}{c^2 r}\bigg)\left[\dfrac{\bigg(\dfrac{GM}{c^2 r}\bigg)\bigg(1-\dfrac{GM}{c^2 r}\bigg)\bigg(3-\dfrac{6\,GM}{c^2 r}-2\bigg(\dfrac{GM}{c^2 r}\bigg)^2\bigg)}{\bigg(1-\dfrac{2GM}{c^2 r}\bigg)^2}+\dfrac{3}{2}\ln{\bigg(1-\dfrac{2GM}{c^2 r}\bigg)}\right]&
\end{flalign*}
Asymptotically (at large $r$), i.e. in the limit $\dfrac{GM}{c^2r}\rightarrow 0$, so that the functions $P_{l'}^{m'}(x)$ and $Q_{l'}^{m'}(x)$ become\,:
\begin{gather}
P_2^2(r\gg R)\approx-3\bigg(\dfrac{c^2 r}{GM}\bigg)^2\quad\text{and}\quad Q_2^2(r\gg R)\approx-\dfrac{8}{5}\bigg(\dfrac{GM}{c^2 r}\bigg)^3\\
\therefore H(r\gg R)\approx-3\,c_P\bigg(\dfrac{c^2 r}{GM}\bigg)^2-\dfrac{8}{5}c_Q \bigg(\dfrac{GM}{c^2 r}\bigg)^3
\end{gather}
From equation (13), $\because$ outside the star,
$e^{2w}=e^{-2\lambda}=\bigg(1-\dfrac{2GM}{c^2 r}\bigg)$,
\begin{gather}
\therefore\Tilde{g}_{tt}(r)=-c^2\bigg(1-\dfrac{2GM}{c^2 r}\bigg)\big(1+H(r)Y_{lm}(\theta,\phi)\big)\nonumber\\
\implies \Tilde{g}_{tt}(r\gg R)\approx-c^2\big(1+H(r\gg R)Y_{lm}(\theta,\phi)\big)
\end{gather}
From equation (16),
\begin{gather}
\Tilde{g}_{tt}(r\gg R)\approx-c^2+\dfrac{3\mathcal{GI}_{ij}}{r^3}n^i n^j-\mathcal{E}_{ij} n^i n^jr^2
\end{gather}
Using equation (17) assuming only one non-vanishing $\mathcal{E}_m$ (and one $\mathcal{I}_m$), equation (37) can be written as:
\begin{gather}
\Tilde{g}_{tt}(r\gg R)\approx-c^2+\dfrac{3}{r^3}\mathcal{I}_{m}Y_{lm}(\theta,\phi)-r^2\mathcal{E}_{m}Y_{lm}(\theta,\phi)
\end{gather}
Using equations (36) and (38),
\begin{gather}
-H(r\gg R)=\dfrac{3G}{c^2r^3}\mathcal{I}-\dfrac{r^2}{c^2}\mathcal{E}
\end{gather}
where $\mathcal{I}=\mathcal{I}_{m}$ and $\mathcal{E}=\mathcal{E}_{m}$. Applying equations (19) and (35) to equation (39),
\begin{gather}
\dfrac{8}{5}c_Q \bigg(\dfrac{GM}{c^2 r}\bigg)^3+3\,c_P\bigg(\dfrac{c^2 r}{GM}\bigg)^2=-\dfrac{3G}{c^2r^3}\eta_2\mathcal{E}-\dfrac{r^2}{c^2}\mathcal{E}\\
\implies c_Q=-\dfrac{15}{8}\bigg(\dfrac{c^2}{G}\bigg)^2\dfrac{1}{M^3}\eta_2\mathcal{E} \quad\text{and}\quad c_P=-\dfrac{1}{3}\bigg(\dfrac{G}{c^2}\bigg)^2\dfrac{M^2}{c^2}\mathcal{E}
\end{gather}
The derivative of equation (33) is
\begin{gather}
H'(x)=c_P {P_2^2}'(x)+c_Q {Q_2^2}'(x)
\end{gather}
Define $y(r)=rH'/H$ and divide equation (42) by equation (33) to get
\begin{gather}
y=\dfrac{c^2r}{GM}\left\{\dfrac{c_P {P_2^2}'(x)+c_Q {Q_2^2}'(x)}{c_P P_2^2(x)+c_Q Q_2^2(x)}\right\}=(1+x)\left\{\dfrac{{P_2^2}'(x)+(c_Q/c_P) {Q_2^2}'(x)}{P_2^2(x)+(c_Q/c_P) Q_2^2(x)}\right\}
\end{gather}
This eliminates $\mathcal{E}$ leaving $\eta_2$ to be expressed in terms of $y_2:=y(r=R)=RH'/H$.
\begin{gather}
y_2=\dfrac{1}{C}\left\{\dfrac{{P_2^2}'(x)+a_2 {Q_2^2}'(x)}{P_2^2(x)+a_2 Q_2^2(x)}\right\}_{x=1/C-1}\\
\implies a_2=-\left\{\dfrac{Cy_2 P_2^2(x)- {P_2^2}'(x)}{Cy_2 Q_2^2(x)- {Q_2^2}'(x)}\right\}_{x=1/C-1}
\end{gather}
where $C=GM/c^2R$ is the compactness parameter (for BH, $C=C_{\text{max}}=1/2$ since $R_\text{BH}=2\,GM/c^2$), and
\begin{gather}
a_2=\dfrac{c_Q}{c_P}=\dfrac{45}{8}\bigg(\dfrac{c^2}{G}\bigg)^4\dfrac{c^2}{M^5}\eta_2=\dfrac{15}{4}C^{-5}k_2\qquad[\text{using eqns. (41) and (21)}]
\end{gather}
From equations (45) and (46), the expression for $k_2$ is obtained as (for convenience, $y\equiv y_2$)
\begin{flalign}
k_2&=\dfrac{8}{5}C^5(1-2C)^2\big[2+2C(y-1)-y\big]\Big\{2C\big[6-3y+3C(5y-8)\big]\nonumber&\\
&+4C^3\big[13-11y+C(3y-2)+2C^2(y+1)\big]+3(1-2C)^2\big[2-y+2C(y-1)\big]\ln{(1-2C)}\Big\}^{-1}&
\end{flalign}
Inside the star, i.e. at $r\leq R$, split equation (22) as follows (prime denotes differentiation w.r.t. $r$):
\begin{gather}
H'=\beta\\
\beta'+C_0 \beta+C_1 H=0\nonumber\\
\implies \beta'=\bigg(1-\dfrac{2Gm}{c^2 r}\bigg)^{-1}\boldsymbol{\beta}\bigg[-\dfrac{2}{r}+\dfrac{G}{c^2}\bigg(\dfrac{2m}{r^2}+4\pi r\Big(\rho-\dfrac{p}{c^2}\Big)\bigg)\bigg]\nonumber\\+\bigg(1-\dfrac{2Gm}{c^2 r}\bigg)^{-1}\boldsymbol{H}\bigg[\dfrac{6}{r^2}-\dfrac{4\pi G}{c^2}\bigg(\psi\Big(\rho+\dfrac{p}{c^2}\Big)+5\rho+\dfrac{9p}{c^2}\bigg)+\dfrac{4G^2}{c^4r^4}\bigg(1-\dfrac{2Gm}{c^2 r}\bigg)^{-1}\bigg(m+\dfrac{4\pi r^3}{c^2}p\bigg)^2\bigg]
\end{gather}
With the (dimensionless) logarithmic derivative, $y=\dfrac{r\beta}{H}=\dfrac{rH'}{H}$, then, it can be written,
\begin{gather}
rH'-yH=0\\
y'H+(y-1)\beta-r\beta'=0\implies y'H+(y-1)\dfrac{yH}{r}-r(\beta C_2+H C_3)=0\nonumber\\
\implies y'+\dfrac{y^2}{r}-\dfrac{y}{r}-y C_2-r C_3=0\implies ry'+y^2-y(1+r C_2)-r^2 C_3=0
\end{gather}
which is a first order non-linear ordinary differential equation with non-constant coefficients, particularly, a Riccati equation, which can be cast back into equation (22). One can either work with equations (47) and (48), or, with equations (49) and (50), (either choice) augmented with equations (23) and (29) to obtain a numerical solution for $H$ using the Runge-Kutta fourth order (RK4) method which is matched with the exterior solution of $H$ ar $r=R$ obtained analytically with two undetermined coefficients.

\subsubsection*{Calculation of axial TLN}
\textbf{Objective}: to obtain expression for the $l=2$ axial tidal Love number, $j_2(N,N')$, and the solution for $N$\\
\textbf{Ingredients:} the form of the metric, perturbation of the metric (in Regge-Wheeler gauge), TOV equations, continuity equation of stress-energy tensor, asymptotic expansion of the metric\\
\textbf{Prescription:} It is convenient to substitute $\cos^{-1}{z}$ for $\theta$ while defining the metric and the perturbation at the beginning. 1. From the perturbation of the metric and the stress-energy tensor, obtain the linearized EFE, $\delta G_{\mu\nu}=\kappa\,\delta T_{\mu\nu}$. 2. From $\delta G_{r\phi}=\kappa\,\delta T_{r\phi}$, realize that $N_1=0$ for stationary perturbations. 3. From $\delta G_{t\phi}=\kappa\,\delta T_{t\phi}$, obtain the (second order) differential equation for $N$ ($N$-equation). 4. Obtain the exterior form of the $N$-equation and get its analytical solution. 5. From the asymptotic behaviour of the exterior solution, determine the coefficients of the exterior solution. 6. Solve for $j_2$ in terms of $N'/N$ at the star's surface $r=R$. 7. For the interior case, split the $N$-equation into two first order differential equations. 8. Solve the coupled differential equations numerically (along with the appropriate TOV equations). 9. Match the interior solution with the exterior solution as well as their first derivatives at $r=R$.\\\\
The analogue of equation (15) in the axial case is
\begin{equation}
\mathcal{A}_{ij}=-\chi \mathcal{B}_{ij}
\end{equation}
where $\mathcal{A}_{ij}$ and $\mathcal{B}_{ij}$ are both ($3 \times 3$) symmetric and traceless (STF: symmetric tracefree) tensors, and are defined to be coefficients in an asymptotic expansion of the total metric at large distances, $r$, from the star.
\begin{gather}
\Tilde{g}_{t\phi}=\Tilde{g}_{\phi t}=-c^2\bigg(1-\dfrac{2GM}{c^2r}\bigg)+\dfrac{3G\mathcal{A}_{ij}}{r^3}n^i n^j-\dfrac{2}{3}\epsilon_{3ki}\mathcal{B}^k_{\;j} n^i n^jr^2+\dots
\end{gather}
with $n^i=x^i/r=(\sin{\theta}\cos{\phi},\sin{\theta}\sin{\phi},\cos{\theta})$ is the unit radial vector, and $\mathcal{B}_{ij}=\dfrac{1}{2}\epsilon_{ipq}\mathcal{R}^{pq}_{\;\;j0}$. Now, $\mathcal{A}_{ij}$ and $\mathcal{B}_{ij}$ can be decomposed as\,:
\begin{gather}
\mathcal{I}_{ij}=\displaystyle\sum_{m=-l}^{l}\mathcal{I}_{m}\mathcal{Y}_{ij}^{lm}\quad \text{and} \quad \mathcal{E}_{ij}=\displaystyle\sum_{m=-l}^{l}\mathcal{E}_{m}\mathcal{Y}_{ij}^{lm}
\end{gather}
where the STF tensors, $\mathcal{Y}_{ij}^{lm}$ are defined as
\begin{gather}
Y_{lm}(\theta,\phi)=\mathcal{Y}_{ij}^{lm}n^in^j
\end{gather}
Thus, equation (15) becomes\,:
\begin{equation}
\mathcal{I}_{m}=-\eta_l \mathcal{E}_{m}
\end{equation}
The (dimensionless) second tidal Love number for degree $l$ is
\begin{equation}
k_l=\dfrac{(2l-1)!!G\eta_l}{2R^{2l+1}}
\end{equation}
so that for $l=$2 (quadrupole), 3 (octapole) and 4 (hexadecupole),
\begin{equation}
k_2=\dfrac{3G\eta_2}{2R^5}\qquad\qquad k_3=\dfrac{15G\eta_3}{2R^7}\qquad\qquad k_4=\dfrac{105G\eta_4}{2R^9}
\end{equation}
For axial perturbations, the metric perturbation in the Regge-Wheeler gauge takes the form:
\begin{equation}
h_{\mu\nu}^\text{axial}=\begin{pmatrix}0 & 0 & N(r)S^{lm}_{\theta}(\theta,\phi) & N(r)S^{lm}_{\phi}(\theta,\phi)\\ 0 & 0 & N_1(r)S^{lm}_{\theta}(\theta,\phi) & N_1(r)S^{lm}_{\phi}(\theta,\phi)\\N(r)S^{lm}_{\theta}(\theta,\phi) & N_1(r)S^{lm}_{\theta}(\theta,\phi) & 0 & 0\\N(r)S^{lm}_{\phi}(\theta,\phi) & N_1(r)S^{lm}_{\phi}(\theta,\phi) & 0 & 0 \end{pmatrix}
\end{equation}
where $S^{lm}_{\theta}(\theta,\phi)=-\partial_{\phi}Y_{lm}(\theta,\phi)/\sin{\theta}$ and $S^{lm}_{\phi}(\theta,\phi)=\sin{\theta}\,\partial_{\theta}Y_{lm}(\theta,\phi)$. It will be shown that $N_1(r)=0$. For $l=2$ and $m=0$, $$S^{lm}_{\theta}(\theta,\phi)=S^{20}_{\theta}(\theta,\phi)=0$$
$$\text{and, } S^{lm}_{\phi}(\theta,\phi)=S^{20}_{\phi}(\theta,\phi)=-\dfrac{3}{2}\sqrt{\dfrac{5}{\pi}}\cos{\theta}(1-\cos^2{\theta})$$
The working metric perturbation is then: $h_{\mu\nu}^\text{axial}=\begin{pmatrix}0 & 0 & 0 & N(r)\\ 0 & 0 & 0 & N_1(r)\\0 & 0 & 0 & 0\\N(r) & N_1(r) & 0 & 0 \end{pmatrix} S^{lm}_{\phi}(\theta,\phi)$\\
Note that total metric perturbation, $\delta g_{\mu\nu}=h_{\mu\nu}^\text{polar}+h_{\mu\nu}^\text{axial}$\\
$$=\begin{pmatrix}-c^2e^{2w}H(r)Y_{20}(\theta,\phi) & 0 & 0 & N(r)S^{20}_{\phi}(\theta,\phi)\\ 0 & e^{2\lambda}H(r)Y_{20}(\theta,\phi) & 0 & N_1(r)S^{20}_{\phi}(\theta,\phi)\\0 & 0 & r^2K(r)Y_{20}(\theta,\phi) & 0\\N(r)S^{20}_{\phi}(\theta,\phi) & N_1(r)S^{20}_{\phi}(\theta,\phi) & 0 & r^2\sin^2{\theta}K(r)Y_{20}(\theta,\phi) \end{pmatrix}$$
The stationary perturbations in stress-energy tensor are given as:
$$T_{\mu\nu} = \left(\rho+\dfrac{p}{c^2}\right)u_{\mu}u_{\nu}+pg_{\mu\nu}\implies \delta T_{\mu\nu} = \left(\rho+\dfrac{p}{c^2}\right)(\delta u_{\mu}u_{\nu}+u_{\mu}\delta u_{\nu})+\left(\delta\rho+\dfrac{\delta p}{c^2}\right)u_{\mu}u_{\nu}+p\,\delta g_{\mu\nu}+\delta p\,g_{\mu\nu}$$
Irrotational case: $\delta u_{\alpha}=0$
$$\therefore \delta T_{\mu\nu} = \left(\delta\rho+\dfrac{\delta p}{c^2}\right)u_{\mu}u_{\nu}+p\,\delta g_{\mu\nu}+\delta p\,g_{\mu\nu}$$
$$\implies \delta T_{t\phi}=\delta T_{\phi t}=pN(r)S^{20}_{\phi}(\theta,\phi)\quad\text{and}\quad\delta T_{r\phi}=\delta T_{\phi r}=pN_1(r)S^{20}_{\phi}(\theta,\phi)$$
Static case: $\delta u^{\alpha}=0$
$$\therefore \delta T_{\mu\nu} = \left(\rho+\dfrac{p}{c^2}\right)(\delta u_{\mu}u_{\nu}+u_{\mu}\delta u_{\nu})+\left(\delta\rho+\dfrac{\delta p}{c^2}\right)u_{\mu}u_{\nu}+p\,\delta g_{\mu\nu}+\delta p\,g_{\mu\nu}$$
$$\delta T_{\mu\nu} = \left(\rho+\dfrac{p}{c^2}\right)(u^{\lambda}\delta g_{\mu\lambda}u_{\nu}+u_{\mu}u^{\lambda}\delta g_{\nu\lambda})+\left(\delta\rho+\dfrac{\delta p}{c^2}\right)u_{\mu}u_{\nu}+p\,\delta g_{\mu\nu}+\delta p\,g_{\mu\nu}$$
$$[\because u_{\sigma}=g^{\lambda\sigma}u_{\lambda}\implies \delta u_{\sigma}=\delta g^{\lambda\sigma}u_{\lambda}+g^{\lambda\sigma}\delta u_{\lambda}\implies \delta u_{\sigma}=u_{\lambda}\delta g^{\lambda\sigma}]$$
$$\implies \delta T_{t\phi}=\left(\rho+\dfrac{p}{c^2}\right)(u^{\lambda}\delta g_{t\lambda}u_{\phi}+u_{t}u^{\lambda}\delta g_{\phi\lambda})+p\,\delta g_{t\phi}=\left(\rho+\dfrac{p}{c^2}\right)u_{t}u^{\lambda}\delta g_{\phi\lambda}+p\,\delta g_{t\phi}$$
$$=\left(\rho+\dfrac{p}{c^2}\right)(-c^2e^w)\,e^{-w}N(r)S^{20}_{\phi}(\theta,\phi)+pN(r)S^{20}_{\phi}(\theta,\phi)=-\rho c2 N(r)S^{20}_{\phi}(\theta,\phi)=\delta T_{\phi t}$$
$$\text{and, }\delta T_{r\phi}=\left(\rho+\dfrac{p}{c^2}\right)(u^{\lambda}\delta g_{r\lambda}u_{\phi}+u_{r}u^{\lambda}\delta g_{\phi\lambda})+p\,\delta g_{r\phi}=p\,\delta g_{t\phi}=pN_1(r)S^{20}_{\phi}(\theta,\phi)=p N_1(r)S^{20}_{\phi}(\theta,\phi)=\delta T_{\phi r}$$
Making use of eqns. (23--30), we thus obtain three equations from the first-order perturbed Einstein's equations $\delta \mathcal{G}_{\mu\nu}=\dfrac{8\pi G}{c^4}\delta T_{\mu\nu}$:\\
\refstepcounter{equation}
\[\resizebox{\linewidth}{!}{$\delta \mathcal{G}_{03}=\dfrac{8\pi G}{c^4}\delta T_{03}\implies e^{-2\lambda}N''-\dfrac{4\pi Gr}{c^4}(p+\rho c^2)N'-\dfrac{1}{r^3}\Big(6r-\dfrac{4Gm}{c^2}-\dfrac{8\pi Gr^3}{c^4}(p+\rho c^2)\Big)N=0\quad[\textbf{irrotational}]$\quad(\theequation)}\]
\refstepcounter{equation}
\[\resizebox{\linewidth}{!}{$\delta \mathcal{G}_{03}=\dfrac{8\pi G}{c^4}\delta T_{03}\implies e^{-2\lambda}N''-\dfrac{4\pi Gr}{c^4}(p+\rho c^2)N'-\dfrac{1}{r^3}\Big(6r-\dfrac{4Gm}{c^2}+\dfrac{8\pi Gr^3}{c^4}(p+\rho c^2)\Big)N=0\quad[\textbf{static}]$\quad(\theequation)}\]
\begin{gather}
\delta \mathcal{G}_{13}=\dfrac{8\pi G}{c^4}\delta T_{13}\implies\dfrac{2N_1}{r^2}=0\implies N_1=0\quad[\textbf{either case}]\\
\delta \mathcal{G}_{23}=\dfrac{8\pi G}{c^4}\delta T_{23}\implies e^{-2\lambda}N_1'+\dfrac{G}{c^2r^2}\Big(2m+4\pi r^3\Big(\dfrac{p}{c^2}-\rho\Big)\Big)N_1=0\quad[\textbf{either case}]
\end{gather}
Equations (53) and (54) can be called the $N$-equation for Irrotational and Static cases respectively analogous to the $H$-equation i.e. equation (22). Notice the change of sign in the last term in the
Outside the star, i.e. at $r\geq R$, $p=\rho=0$ and $m=M$, so that $e^{-2\lambda}=\bigg(1-\dfrac{2GM}{c^2 r}\bigg)$. The $N$-equation then becomes
\begin{gather}
\bigg(1-\dfrac{2GM}{c^2 r}\bigg)N''-\dfrac{1}{r^2}\Big(6-\dfrac{4GM}{c^2r}\Big)N=0\quad[\textbf{either case}]
\end{gather}
Now, put $x=\bigg(\dfrac{c^2 r}{GM}-1\bigg)$ i.e. $r=\dfrac{GM}{c^2}(x+1)$ so that $\dfrac{d}{dr}=\dfrac{dx}{dr}\dfrac{d}{dx}=\bigg(\dfrac{c^2}{GM}\bigg)\dfrac{d}{dx}$ and $\dfrac{d^2}{dr^2}=\bigg(\dfrac{c^2}{GM}\bigg)^2\dfrac{d^2}{dx^2}$
\begin{gather}
\therefore \bigg(\dfrac{x-1}{x+1}\bigg)N''-\dfrac{1}{(x+1)^2}\bigg(\dfrac{6x+2}{x+1}\bigg)N=0\implies (x^2-1)N''-\bigg(\dfrac{6x+2}{x+1}\bigg)N=0
\end{gather}
where prime denotes differentiation w.r.t $x$. The solution of above equation is given as
\begin{gather}
N(x)=c_AA(x)+c_BB(x)
\end{gather}
where $A(x)=(x-1)(x+1)^2=\bigg(\dfrac{c^2 r}{GM}\bigg)^2\bigg(\dfrac{c^2 r}{GM}-2\bigg)$\\
and $B(x)=\dfrac{1}{(x+1)}(6x^3+12x^2+2x-8)+3(x-1)(x+1)^2\ln{\dfrac{x-1}{x+1}}$\\
$\phantom{}\qquad\qquad=\bigg(\dfrac{GM}{c^2r}\bigg)\bigg(6\bigg(\dfrac{c^2 r}{GM}\bigg)^2\bigg(\dfrac{c^2 r}{GM}-1\bigg)-4\bigg(\dfrac{c^2 r}{GM}\bigg)-4\bigg)+3\bigg(\dfrac{c^2 r}{GM}\bigg)^2\bigg(\dfrac{c^2 r}{GM}-2\bigg)\ln\bigg(1-\dfrac{2GM}{c^2 r}\bigg)$\\
Asymptotically (at large $r$), i.e. in the limit $\dfrac{GM}{c^2r}\rightarrow 0$, so that the functions $A(x)$ and $B(x)$ become\,:
\begin{gather}
A(r\gg R)\approx\bigg(\dfrac{c^2 r}{GM}\bigg)^3\quad\text{and}\quad B(r\gg R)\approx \dfrac{24}{5}\bigg(\dfrac{GM}{c^2 r}\bigg)^2\\
\therefore N(r\gg R)\approx\,c_A\bigg(\dfrac{c^2r}{GM}\bigg)^3+\dfrac{24}{5}c_B \bigg(\dfrac{GM}{c^2 r}\bigg)^2
\end{gather}


\section*{Appendix}
\subsection*{Units}
SI unit of Gravitational constant, $G$:   m$^3$ kg$^{-1}$ s$^{-2}$\\
SI unit of Speed of light, $c$:           m s$^{-1}$\\
SI unit of $\kappa=\dfrac{8\pi G}{c^4}$:  m$^{-1}$ kg$^{-1}$ s$^2$\\
SI unit of Pressure, $p$:                 m$^{-1}$ kg s$^{-2}$\\
SI unit of Energy density, $\varepsilon$: m$^{-1}$ kg s$^{-2}$\\
SI unit of Matter density, $\rho$:        m$^{-3}$ kg\\
SI unit of Ricci scalar, $R$:             m$^{-2}$\\
SI unit of Tidal field, $\mathcal{E}\sim\dfrac{GM}{r^3}$: s$^{-2}$\\
SI unit of Mass quadrupole moment, $Q$: m$^2$ kg\\
SI unit of Tidal deformability, $\eta$: m$^2$ kg s$^2$\\\\
1 MeV fm$^{-3} = 1.6022\times 10^{32}$ J m$^{-3} = 1.6022\times 10^{33}$ erg cm$^{-3}$\\
1 MeV fm$^{-3} \equiv 1.7827\times 10^{15}$ kg m$^{-3} = 1.7827\times 10^{12}$ g cm$^{-3}$

\subsection*{Stress-Energy-Momentum Tensor}
Coordinates: $(t,r,\theta, \phi)$ \qquad [\,Alternative choices are $(ct,r,\theta, \phi)$ and $(ict,r,\theta, \phi)$\,] \\
Signature: $(\,-\,+\,+\,+\,)$ \qquad [\,Alternative choices are $(\,+\,-\,-\,-\,)$ and $(\,+\,+\,+\,+\,)$\,] \\
Metric for spherical symmetry, $g_{\mu\nu}$ = $g_{\nu\mu}$ = $\begin{pmatrix}-c^2 e^{2w} & 0 & 0 & 0\\ 0 & e^{2\lambda} & 0 & 0\\0 & 0 & r^2 & 0\\0 & 0 & 0 & r^2 \sin^2{\theta} \end{pmatrix}$\\
Line element, $ds=\sqrt{-g_{\mu\nu}dx^{\mu} dx^{\nu}}$\\
Proper time, $d\tau=\dfrac{ds}{c}$\\
Inverse metric, $g^{\mu\nu}$ = $g^{\nu\mu}$ = $\begin{pmatrix}-c^{-2} e^{-2w} & 0 & 0 & 0\\ 0 & e^{-2\lambda} & 0 & 0\\0 & 0 & r^{-2} & 0\\0 & 0 & 0 & r^{-2} \csc^2{\theta} \end{pmatrix}$\\
Trace of metric, $\text{Tr}(g_{\mu\nu})=g^{\mu\nu}g_{\mu\nu} = 4$\\
Four velocity of static fluid, $u^0=\dfrac{c}{\sqrt{-g_{00}}}$ and $u^i=0$ so that $u^{\mu}$ = $\begin{pmatrix}e^{-w} \\ 0 \\ 0 \\ 0 \end{pmatrix}$\\
$u^{\mu}u^{\nu}$ = $\begin{pmatrix}e^{-2w} & 0 & 0 & 0\\ 0 & 0 & 0 & 0 \\ 0 & 0 & 0 & 0 \\ 0 & 0 & 0 & 0 \end{pmatrix}\implies u^{\mu}u^{\nu}g_{\mu\nu}=-c^2\implies u^{\mu}u_{\mu}=-c^2\implies \delta(u_{\mu}u^{\mu})=0$\\
$u_{\nu}$ = $g_{\nu\mu}u^{\mu}$ = $g_{\mu\nu}u^{\mu}$ = $\begin{pmatrix}-c^2e^{w} & 0 & 0 & 0 \end{pmatrix}\implies u_{\mu}u_{\nu} = \begin{pmatrix}c^4e^{2w} & 0 & 0 & 0\\ 0 & 0 & 0 & 0\\0 & 0 & 0 & 0\\0 & 0 & 0 & 0 \end{pmatrix}\implies u_{\mu}u_{\nu}g^{\mu\nu}=-c^2$\\
$u_{\mu}u^{\nu}=\begin{pmatrix} -c^2 & 0 & 0 & 0\\ 0 & 0 & 0 & 0\\0 & 0 & 0 & 0\\0 & 0 & 0 & 0\end{pmatrix}\implies\delta(u_{\mu}u^{\nu})=0_{4,4}$\\
Stress–energy–momentum tensor, $T^{\mu\nu} = \left(\rho+\dfrac{p}{c^2}\right)u^{\mu}u^{\nu}+\,pg^{\mu\nu}=\begin{pmatrix} e^{-2w}\rho & 0 & 0 & 0\\ 0 & e^{-2\lambda}p & 0 & 0\\0 & 0 & r^{-2}p & 0\\0 & 0 & 0 & r^{-2}p \csc^2{\theta}\end{pmatrix}$\\
Alternatively, $T^{\mu\nu} = \left(\dfrac{\varepsilon}{c^2}+\dfrac{p}{c^2}\right)u^{\mu}u^{\nu}+\,pg^{\mu\nu}=\begin{pmatrix} e^{-2w}\varepsilon c^{-2} & 0 & 0 & 0\\ 0 & e^{-2\lambda}p & 0 & 0\\0 & 0 & r^{-2}p & 0\\0 & 0 & 0 & r^{-2}p \csc^2{\theta}\end{pmatrix}$\\
$T_{\mu\nu}=g_{\mu\alpha}g_{\nu\beta}T^{\alpha\beta}=\begin{pmatrix} e^{2w}\rho c^4 & 0 & 0 & 0\\ 0 & e^{2\lambda}p & 0 & 0\\0 & 0 & r^2 p & 0\\0 & 0 & 0 & r^2 p\sin^2{\theta} \end{pmatrix}=\begin{pmatrix} e^{2w}\varepsilon c^2 & 0 & 0 & 0\\ 0 & e^{2\lambda}p & 0 & 0\\0 & 0 & r^2 p & 0\\0 & 0 & 0 & r^2 p\sin^2{\theta} \end{pmatrix}$\\
Also, $T_{\mu\nu} = \left(\rho+\dfrac{p}{c^2}\right)u_{\mu}u_{\nu}+pg_{\mu\nu}=\begin{pmatrix} e^{2w}\rho c^4 & 0 & 0 & 0\\ 0 & e^{2\lambda}p & 0 & 0\\0 & 0 & r^2 p & 0\\0 & 0 & 0 & r^2 p\sin^2{\theta} \end{pmatrix}$\\
Alternatively, $T_{\mu\nu} = \left(\dfrac{\varepsilon}{c^2}+\dfrac{p}{c^2}\right)u_{\mu}u_{\nu}+\,pg_{\mu\nu}=\begin{pmatrix} e^{2w}\varepsilon c^2 & 0 & 0 & 0\\ 0 & e^{2\lambda}p & 0 & 0\\0 & 0 & r^2 p & 0\\0 & 0 & 0 & r^2p \sin^2{\theta}\end{pmatrix}$\\
$T^{\,\,\nu}_{\mu}=g^{\mu\sigma}T_{\sigma\nu}=\begin{pmatrix}-c^{-2} e^{-2w} & 0 & 0 & 0\\ 0 & e^{-2\lambda} & 0 & 0\\0 & 0 & r^{-2} & 0\\0 & 0 & 0 & r^{-2} \csc^2{\theta} \end{pmatrix}\begin{pmatrix} e^{2w}\rho c^4 & 0 & 0 & 0\\ 0 & e^{2\lambda}p & 0 & 0\\0 & 0 & r^2 p & 0\\0 & 0 & 0 & r^2 p\sin^2{\theta} \end{pmatrix}=\begin{pmatrix} -\rho c^2 & 0 & 0 & 0\\ 0 & p & 0 & 0\\0 & 0 & p & 0\\0 & 0 & 0 & p \end{pmatrix}$\\
$T^{\,\,\nu}_{\mu}=\left(\rho+\dfrac{p}{c^2}\right)u_{\mu}u^{\nu}+p\,\delta^{\,\,\nu}_{\mu}=\left(\rho+\dfrac{p}{c^2}\right) \begin{pmatrix} -c^2 & 0 & 0 & 0\\ 0 & 0 & 0 & 0\\0 & 0 & 0 & 0\\0 & 0 & 0 & 0\end{pmatrix}+p \begin{pmatrix} 1 & 0 & 0 & 0\\ 0 & 1 & 0 & 0\\0 & 0 & 1 & 0\\0 & 0 & 0 & 1\end{pmatrix}=\begin{pmatrix} -\rho c^2 & 0 & 0 & 0\\ 0 & p & 0 & 0\\0 & 0 & p & 0\\0 & 0 & 0 & p \end{pmatrix}$
\vspace{.5em}
\\Trace of stress–energy–momentum tensor, $\text{Tr}\,(T_{\mu\nu}) = g_{\mu\nu}T^{\mu\nu} = g^{\mu\nu}T_{\mu\nu} = \text{Tr}\,(T^{\mu\nu}) = 3p-\rho c^2$
\vspace{.5em}
\\Now, $u_{\mu}u^{\mu}=g^{\mu\nu}u_{\mu}u_{\nu}\implies \delta g^{\mu\nu}u_{\mu}u_{\nu}+ g^{\mu\nu}\delta (u_{\mu}u_{\nu})=0\quad[\,\because \delta (u_{\mu}u^{\mu})=0\,]\implies  g^{\mu\nu}\delta (u_{\mu}u_{\nu})=-\delta g^{\mu\nu}u_{\mu}u_{\nu}$\\\\
$\implies g^{\mu\nu}(u_{\nu}\delta u_{\mu}+ u_{\mu}\delta u_{\nu})=-\delta g^{\mu\nu}u_{\mu}u_{\nu}\implies u^{\mu}\delta u_{\mu}+ u^{\nu}\delta u_{\nu}=-\delta g^{\mu\nu}u_{\mu}u_{\nu}\implies u^{\mu}\delta u_{\mu}=-\dfrac{1}{2}\delta g^{\mu\nu}u_{\mu}u_{\nu}$\\\\
$\implies u^{\mu}\delta u_{\mu}=-\dfrac{1}{2}(-g^{\mu\alpha}g^{\nu\beta}\delta g_{\alpha\beta})u_{\mu}u_{\nu}\implies u^{\mu}\delta u_{\mu}=\dfrac{1}{2}\delta g_{\alpha\beta}u^{\alpha}u^{\beta}$\\\\
Then, the time component is $e^{-w}\delta u_t=\dfrac{1}{2}\delta g_{tt}e^{-2w}\implies \delta u_t=\dfrac{1}{2}e^{-w}\delta g_{tt}$

\subsection*{Hydrostatic Equilibrium}
\begin{flalign*}
&\because \nabla_{\boldsymbol{l}} C^i_{jk}=\partial_{\boldsymbol{l}} C^i_{jk}+\Gamma^i_{\boldsymbol{l}m}C^m_{jk}-\Gamma^m_{\boldsymbol{l}j}C^i_{mk}-\Gamma^m_{\boldsymbol{l}k}C^i_{jm}&\\
&\therefore \nabla_{\boldsymbol{\mu}}T^{\mu\nu}=\partial_{\boldsymbol{\mu}} T^{\mu\nu}+\Gamma^{\mu}_{\boldsymbol{\mu}m}T^{m\nu}+\Gamma^{\nu}_{\boldsymbol{\mu}m}T^{\mu m}&\\\\
&\nabla_{\boldsymbol{\mu}}T^{\mu 0}=\partial_{\boldsymbol{\mu}} T^{\mu 0}+\Gamma^{\mu}_{\boldsymbol{\mu}m}T^{m 0}+\Gamma^{0}_{\boldsymbol{\mu}m}T^{\mu m}&\\
&=\partial_{\mu} T^{\mu 0}+\left(\Gamma^{\mu}_{\mu 0}T^{0 0}\right)+\left(\Gamma^{0}_{\mu 0}T^{\mu 0}+\Gamma^{0}_{\mu 1}T^{\mu 1}+\Gamma^{0}_{\mu 2}T^{\mu 2}+\Gamma^{0}_{\mu 3}T^{\mu 3}\right)&\\
&=\partial_{0} T^{00}+\left(\Gamma^{0}_{00}T^{00}+\Gamma^{1}_{10}T^{00}+\Gamma^{2}_{20}T^{00}+\Gamma^{3}_{30}T^{00}\right)+\left(\Gamma^{0}_{00}T^{00}+\Gamma^{0}_{11}T^{11}+\Gamma^{0}_{22}T^{22}+\Gamma^{0}_{33}T^{33}\right)=0&\\\\
&\nabla_{\boldsymbol{\mu}}T^{\mu 1}=\partial_{\boldsymbol{\mu}} T^{\mu 1}+\Gamma^{\mu}_{\boldsymbol{\mu}m}T^{m 1}+\Gamma^{1}_{\boldsymbol{\mu}m}T^{\mu m}&\\
&=\partial_{\mu} T^{\mu 1}+\left(\Gamma^{\mu}_{\mu 1}T^{11}\right)+\left(\Gamma^{1}_{\mu 0}T^{\mu 0}+\Gamma^{1}_{\mu 1}T^{\mu 1}+\Gamma^{1}_{\mu 2}T^{\mu 2}+\Gamma^{1}_{\mu 3}T^{\mu 3}\right)&\\
&=\partial_{1} T^{11}+\left(\Gamma^{0}_{01}T^{11}+\Gamma^{1}_{11}T^{11}+\Gamma^{2}_{21}T^{11}+\Gamma^{3}_{31}T^{11}\right)+\left(\Gamma^{1}_{00}T^{00}+\Gamma^{1}_{11}T^{11}+\Gamma^{1}_{22}T^{22}+\Gamma^{1}_{33}T^{33}\right)&\\
&=\partial_{1} T^{11}+\left(w'T^{11}+\lambda'T^{11}+\dfrac{1}{r}T^{11}+\dfrac{1}{r}T^{11}\right)+\left(c^2 e^{2w-2\lambda}w'T^{00}+\lambda'T^{11}-e^{-2\lambda}rT^{22}-e^{-2\lambda}r\sin^2{\theta}T^{33}\right)&\\
&=\dfrac{dp}{dr}e^{-2\lambda}-2\lambda'e^{-2\lambda}p+\left(w'+\lambda'+\dfrac{1}{r}+\dfrac{1}{r}\right)e^{-2\lambda}p+\left(c^2 e^{-2\lambda}w'\rho+\lambda'e^{-2\lambda}p-e^{-2\lambda}r^{-1}p-e^{-2\lambda}r^{-1}p\right)&\\
&=\dfrac{dp}{dr}e^{-2\lambda}+w'e^{-2\lambda}p+c^2 e^{-2\lambda}w'\rho&
\end{flalign*}
Continuity equation of the energy-momentum tensor yields the hydrostatic equilibrium condition: $$\nabla_{\boldsymbol{\mu}}T^{\mu\nu}=0\implies\nabla_{\boldsymbol{\mu}}T^{\mu 1}=0\implies\dfrac{dp}{dr}+w'p+c^2 w'\rho=0\implies\dfrac{dp}{dr}=-(\rho c^2+p)\dfrac{dw}{dr}$$
\begin{flalign*}
\text{From TOV equations, } \dfrac{dw}{dr}&=\dfrac{1}{r}(1-2\,C_r)^{-1}\left(C_r+\dfrac{4\pi G }{c^4}r^2p\right)\quad \left[\text{where } C_r(r)=\dfrac{Gm(r)}{c^2r}\right]&\\
&\approx \dfrac{Gm}{c^2r^2}+\dfrac{4\pi G }{c^4}rp\quad\left[\text{when }2\,C_r\ll 1\right]\quad=\dfrac{G}{c^2r^2}\left(m+\dfrac{4\pi r^3}{c^2}p\right)&
\end{flalign*}
\subsection*{TOV equations in GR}
\textbf{Objective}: to obtain equations for pressure gradient (equation of hydrostatic equilibrium in GR) and for the metric parameters, $w$ and $\lambda$\\
\textbf{Ingredients:} the form of the metric, Einstein's field equations (EFE), continuity equation of stress-energy tensor (optional)\\
\textbf{Prescription:} 1. From the $tt$-component of EFE, get the $\lambda$-equation. 2. From the $rr$-component, get the $w$-equation. 3. Use the $r$-component of stress-energy continuity equation and the $w$-equation to get equation of hydrostatic equilibrium. Alternatively, use the $tt$-component and $rr$-component to eliminate $\lambda'$, $w'$, $w''$ from the $\theta\theta$-equation (this is the longer route). The validity of this alternative way owes to the continuity (or, conservation) of Einstein tensor (in concert with the continuity of its counterpart, the stress-energy tensor) in the development of the field equations. Start from EFE (mixed form):
$$R_{\mu}^{\,\,\nu}-\dfrac{1}{2}R\delta_{\mu}^{\,\,\nu}=\dfrac{8\pi G}{c^4}T_{\mu}^{\,\,\nu}\implies \mathcal{G}_{\mu}^{\,\,\nu}=\dfrac{8\pi G}{c^4}T_{\mu}^{\,\,\nu}$$
The first three components (the fourth being same as the third) of EFE are:
$$\mathcal{G}_{0}^{\,\,0}=\dfrac{8\pi G}{c^4}T_{0}^{\,\,0}\implies-\dfrac{1}{r^2}e^{-2\lambda}(-1+e^{2\lambda}+2r\lambda')=-\dfrac{8\pi G}{c^4}\rho c^2$$
$$\mathcal{G}_{1}^{\,\,1}=\dfrac{8\pi G}{c^4}T_{1}^{\,\,1}\implies\dfrac{1}{r^2}(-1+e^{-2\lambda}(1+2rw'))=\dfrac{8\pi G}{c^4}p$$
$$\mathcal{G}_{2}^{\,\,2}=\dfrac{8\pi G}{c^4}T_{2}^{\,\,2}\implies\dfrac{1}{r}e^{-2\lambda}((1+rw')(w'-\lambda')+rw'')=\dfrac{8\pi G}{c^4}p$$
The first equation may be written as:
$$e^{-2\lambda}(2r\lambda'-1)+1=\dfrac{8\pi G}{c^2}\rho r^2\implies \dfrac{d}{dr}(r(1-e^{-2\lambda})=\dfrac{8\pi G}{c^2}\rho r^2$$
$$\implies r(1-e^{-2\lambda})=\dfrac{2G}{c^2}\displaystyle\int_0^r 4\pi r^2  \rho\,dr=\dfrac{2Gm}{c^2}\implies e^{-2\lambda}=1-\dfrac{2Gm}{c^2 r}$$
where $m(r)=\displaystyle\int_0^r 4\pi r^2 \rho(r)\,dr$. The second equation then becomes:
$$2rw'=e^{2\lambda}\bigg(1+\dfrac{8\pi G}{c^4}r^2 p\bigg)-1\implies w'=\dfrac{1}{2r}e^{2\lambda}\bigg(1+\dfrac{8\pi G}{c^4}r^2 p-e^{-2\lambda}\bigg)$$
$$\implies w'=\dfrac{1}{2r}\bigg(1-\dfrac{2Gm}{c^2 r}\bigg)^{-1}\bigg(\dfrac{8\pi G}{c^4}r^2 p+\dfrac{2Gm}{c^2 r}\bigg)\implies w'=\dfrac{G}{c^2r^2}\bigg(1-\dfrac{2Gm}{c^2 r}\bigg)^{-1}\bigg(m+\dfrac{4\pi r^3}{c^2}p\bigg)$$
Now, the $r$-component of the continuity equation is $\dfrac{dp}{dr}=-(\rho c^2+p)\dfrac{dw}{dr}$
$$\therefore\,\dfrac{dp}{dr}=-\dfrac{G}{r^2}\bigg(\rho +\dfrac{p}{c^2}\bigg)\bigg(1-\dfrac{2Gm}{c^2 r}\bigg)^{-1}\bigg(m+\dfrac{4\pi r^3}{c^2}p\bigg)$$
The above equation is the equation of hydrostatic equilibrium in GR (for static fluid in spherical symmetry).

\subsection*{Deriving the geodesic equation from the action principle}
Let's try to find a geodesic between two timelike-separated events. (Note that $\mu$ and $\nu$ are dummy indices.)
\begin{flalign*}
&\delta \mathcal{S}=0\implies 0=\delta\displaystyle\int ds=\delta\displaystyle\int \sqrt{-g_{\mu\nu}dx^{\mu}dx^{\nu}}= \delta\displaystyle\int \sqrt{-g_{\mu\nu}\dfrac{dx^{\mu}}{d\tau}\dfrac{dx^{\nu}}{d\tau}}d\tau=\displaystyle\int \delta\bigg(\sqrt{-g_{\mu\nu}\dfrac{dx^{\mu}}{d\tau}\dfrac{dx^{\nu}}{d\tau}}\bigg)d\tau&\\
&=\displaystyle\int \dfrac{ \delta\bigg(-g_{\mu\nu}\dfrac{dx^{\mu}}{d\tau}\dfrac{dx^{\nu}}{d\tau}\bigg)}{2\sqrt{-g_{\mu\nu}\dfrac{dx^{\mu}}{d\tau}\dfrac{dx^{\nu}}{d\tau}}} d\tau=\displaystyle\int \dfrac{ \delta\bigg(-g_{\mu\nu}\dfrac{dx^{\mu}}{d\tau}\dfrac{dx^{\nu}}{d\tau}\bigg)}{2\,\dfrac{ds}{d\tau}} d\tau=\dfrac{1}{2c}\displaystyle\int \delta\bigg(-g_{\mu\nu}\dfrac{dx^{\mu}}{d\tau}\dfrac{dx^{\nu}}{d\tau}\bigg) d\tau&\\
&\implies 0=\displaystyle\int \delta\bigg(g_{\mu\nu}\dfrac{dx^{\mu}}{d\tau}\dfrac{dx^{\nu}}{d\tau}\bigg) d\tau=\displaystyle\int \bigg(\delta g_{\mu\nu}\dfrac{dx^{\mu}}{d\tau}\dfrac{dx^{\nu}}{d\tau}+g_{\mu\nu}\dfrac{d\,\delta x^{\mu}}{d\tau}\dfrac{dx^{\nu}}{d\tau}+g_{\mu\nu}\dfrac{dx^{\mu}}{d\tau}\dfrac{d\,\delta x^{\nu}}{d\tau}\bigg) d\tau\quad[\text{by product rule}]&\\
&=\displaystyle\int \bigg(\dfrac{dx^{\mu}}{d\tau}\dfrac{dx^{\nu}}{d\tau}\partial_{\lambda}g_{\mu\nu}\delta x^{\lambda}+2\,g_{\mu\nu}\dfrac{dx^{\mu}}{d\tau}\dfrac{d\,\delta x^{\nu}}{d\tau}\bigg) d\tau\quad[\because g_{\mu\nu}\equiv g_{\mu\nu}(x^{\lambda})\implies \delta g_{\mu\nu}=\partial_{\lambda}g_{\mu\nu}\delta x^{\lambda}]&\\
&=\displaystyle\int \dfrac{dx^{\mu}}{d\tau}\dfrac{dx^{\nu}}{d\tau}(\partial_{\lambda}g_{\mu\nu})\delta x^{\lambda}d\tau+2\bigg\{g_{\mu\nu}\dfrac{dx^{\mu}}{d\tau}\delta x^{\nu}-\displaystyle\int\dfrac{d}{d\tau}\bigg(g_{\mu\nu}\dfrac{dx^{\mu}}{d\tau}\bigg)\delta x^{\nu}d\tau\bigg\}\quad[\text{integrating by parts}]&\\
&=\displaystyle\int \bigg\{\dfrac{dx^{\mu}}{d\tau}\dfrac{dx^{\nu}}{d\tau}(\partial_{\lambda}g_{\mu\nu})\delta x^{\lambda}-2\dfrac{d}{d\tau}\bigg(g_{\mu\nu}\dfrac{dx^{\mu}}{d\tau}\bigg)\delta x^{\nu}\bigg\}d\tau\quad[\text{the middle term was zero at the boundaries}]&\\
&=\displaystyle\int \bigg\{\dfrac{dx^{\mu}}{d\tau}\dfrac{dx^{\nu}}{d\tau}(\partial_{\lambda}g_{\mu\nu})\delta x^{\lambda}-2g_{\mu\nu}\dfrac{d^2x^{\mu}}{d\tau^2}\delta x^{\nu}-2\partial_{\lambda}g_{\mu\nu}\dfrac{d x^{\lambda}}{d\tau}\dfrac{dx^{\mu}}{d\tau}\delta x^{\nu}\bigg\}d\tau\quad[\text{by product rule, and, }g_{\mu\nu}\equiv g_{\mu\nu}(x^{\lambda})]&\\
&=\displaystyle\int \bigg\{\dfrac{dx^{\mu}}{d\tau}\dfrac{dx^{\lambda}}{d\tau}\partial_{\nu}g_{\mu\lambda}-2g_{\mu\nu}\dfrac{d^2x^{\mu}}{d\tau^2}-2\partial_{\lambda}g_{\mu\nu}\dfrac{d x^{\lambda}}{d\tau}\dfrac{dx^{\mu}}{d\tau}\bigg\}\delta x^{\nu}d\tau\quad[\text{interchanging dummy indices in the first term}]&\\
&=\displaystyle\int \bigg\{-2g_{\mu\nu}\dfrac{d^2x^{\mu}}{d\tau^2}+\dfrac{dx^{\mu}}{d\tau}\dfrac{dx^{\lambda}}{d\tau}\partial_{\nu}g_{\mu\lambda}-\partial_{\lambda}g_{\mu\nu}\dfrac{d x^{\lambda}}{d\tau}\dfrac{dx^{\mu}}{d\tau}-\partial_{\mu}g_{\lambda\nu}\dfrac{d x^{\mu}}{d\tau}\dfrac{dx^{\lambda}}{d\tau}\bigg\}\delta x^{\nu}d\tau&\\
&=\displaystyle\int \bigg\{g_{\mu\nu}\dfrac{d^2x^{\mu}}{d\tau^2}+\dfrac{1}{2}\dfrac{dx^{\mu}}{d\tau}\dfrac{dx^{\lambda}}{d\tau}(-\partial_{\nu}g_{\mu\lambda}+\partial_{\lambda}g_{\mu\nu}+\partial_{\mu}g_{\lambda\nu})\bigg\}\delta x^{\nu}d\tau\quad[\text{dividing throughout by }-2]&\\
&\implies g_{\mu\nu}\dfrac{d^2x^{\mu}}{d\tau^2}+\dfrac{1}{2}\dfrac{dx^{\mu}}{d\tau}\dfrac{dx^{\lambda}}{d\tau}(-\partial_{\nu}g_{\mu\lambda}+\partial_{\lambda}g_{\mu\nu}+\partial_{\mu}g_{\lambda\nu})=0\quad[\text{Euler-Lagrange equation}]&\\
&\implies \delta_{\;\mu}^{\rho}\dfrac{d^2x^{\mu}}{d\tau^2}+\dfrac{1}{2}g^{\rho\nu}\dfrac{dx^{\mu}}{d\tau}\dfrac{dx^{\lambda}}{d\tau}(-\partial_{\nu}g_{\mu\lambda}+\partial_{\lambda}g_{\mu\nu}+\partial_{\mu}g_{\lambda\nu})=0\quad[\text{contracting with }g^{\rho\nu}]&\\
&\implies \boldsymbol{\dfrac{d^2x^{\rho}}{d\tau^2}=-\dfrac{dx^{\mu}}{d\tau}\dfrac{dx^{\lambda}}{d\tau}\Gamma^{\rho}_{\;\mu\lambda}}\quad\big[\text{where }\Gamma^{\rho}_{\;\mu\lambda}=\dfrac{1}{2}g^{\rho\nu}(\partial_{\mu}g_{\lambda\nu}+\partial_{\lambda}g_{\mu\nu}-\partial_{\nu}g_{\mu\lambda}\big]&
\end{flalign*}

\subsection*{Variation of $f(\mathcal{R})$ action}
$$\mathcal{S}=\displaystyle\int d^4x\sqrt{-g}\,\left[\dfrac{1}{2\kappa}f(\mathcal{R})+\mathcal{L}_{\text{M}}\right]$$
According to the action principle, the variation of this action w.r.t. to the \textbf{inverse metric} is zero.
\begin{flalign*}
&\text{i.e. }\delta \mathcal{S}=0\implies \delta\displaystyle\int \mathcal{L}\sqrt{-g}\,d^4x=0\quad[\,\text{where }\mathcal{L}\text{ is the Lagrangian density}\,]&\\
&\implies \delta\displaystyle\int d^4x\sqrt{-g}\,\left[\dfrac{1}{2\kappa}f(\mathcal{R})+\mathcal{L}_{\text{M}}\right]=0&\\
&\implies \displaystyle\int d^4x\,\left[\dfrac{1}{2\kappa}\dfrac{\delta\left(\sqrt{-g}f(\mathcal{R})\right)}{\delta g^{\mu\nu}}+\dfrac{\delta\left(\sqrt{-g}\mathcal{L}_{\text{M}}\right)}{\delta g^{\mu\nu}}\right]\delta g^{\mu\nu}=0&\\
&\implies \displaystyle\int d^4x\,\left[\dfrac{1}{2\kappa}\left(\sqrt{-g}\,\dfrac{\delta f(\mathcal{R})}{\delta g^{\mu\nu}}+f(\mathcal{R})\dfrac{\delta\sqrt{-g}}{\delta g^{\mu\nu}}\right)+\dfrac{\delta\left(\sqrt{-g}\mathcal{L}_{\text{M}}\right)}{\delta g^{\mu\nu}}\right]\delta g^{\mu\nu}=0&\\
&\implies \displaystyle\int d^4x\,\left[\dfrac{1}{2\kappa}\left(\dfrac{\delta f(\mathcal{R})}{\delta g^{\mu\nu}}+\dfrac{f(\mathcal{R})}{\sqrt{-g}}\dfrac{\delta\sqrt{-g}}{\delta g^{\mu\nu}}\right)+\dfrac{1}{\sqrt{-g}}\dfrac{\delta\left(\sqrt{-g}\mathcal{L}_{\text{M}}\right)}{\delta g^{\mu\nu}}\right]\sqrt{-g}\,\delta g^{\mu\nu}=0&\\
&\implies \dfrac{1}{2\kappa}\left(\dfrac{\delta f(\mathcal{R})}{\delta g^{\mu\nu}}+\dfrac{f(\mathcal{R})}{\sqrt{-g}}\dfrac{\delta\sqrt{-g}}{\delta g^{\mu\nu}}\right)+\dfrac{1}{\sqrt{-g}}\dfrac{\delta\left(\sqrt{-g}\mathcal{L}_{\text{M}}\right)}{\delta g^{\mu\nu}}=0&\\
&\implies \dfrac{\delta f(\mathcal{R})}{\delta g^{\mu\nu}}+\dfrac{f(\mathcal{R})}{\sqrt{-g}}\dfrac{\delta\sqrt{-g}}{\delta g^{\mu\nu}}=-2\kappa\dfrac{1}{\sqrt{-g}}\dfrac{\delta\left(\sqrt{-g}\mathcal{L}_{\text{M}}\right)}{\delta g^{\mu\nu}}&\\
&\implies \dfrac{\delta f(\mathcal{R})}{\delta g^{\mu\nu}}-\dfrac{1}{2}f(\mathcal{R})g_{\mu\nu}=\kappa T_{\mu\nu}&
\end{flalign*}
Now, $T_{\mu\nu}:=\dfrac{-2}{\sqrt{-g}}\dfrac{\delta\left(\sqrt{-g}\mathcal{L}_{\text{M}}\right)}{\delta g^{\mu\nu}}$\\
By Jacobi's formula, $\delta g=\delta \det(g_{\mu\nu})=gg^{\mu\nu}\delta g_{\mu\nu}$\\
$\therefore \delta\sqrt{-g}=\dfrac{-1}{2\sqrt{-g}}\delta g=\dfrac{-1}{2\sqrt{-g}}\left(gg^{\mu\nu}\delta g_{\mu\nu}\right)=\dfrac{1}{2}\sqrt{-g}\left(g^{\mu\nu}\delta g_{\mu\nu}\right)=-\dfrac{1}{2}\sqrt{-g}\left(g_{\mu\nu}\delta g^{\mu\nu}\right)$\\
$\implies\dfrac{1}{\sqrt{-g}}\dfrac{\delta\sqrt{-g}}{\delta g^{\mu\nu}}=-\dfrac{1}{2}g_{\mu\nu}$\\
Further, $\dfrac{\delta f(\mathcal{R})}{\delta g^{\mu\nu}}=\dfrac{df(\mathcal{R})}{d\mathcal{R}}\dfrac{\delta\mathcal{R}}{\delta g^{\mu\nu}}= \dfrac{\delta\mathcal{R}}{\delta g^{\mu\nu}}f_{\mathcal{R}}$
\begin{flalign*}
&\text{Riemann curvature tensor, } \mathcal{R}^{\rho}_{\eta\sigma\chi}=\partial_{\sigma}\Gamma^{\rho}_{\chi\eta}-\partial_{\chi}\Gamma^{\rho}_{\sigma\eta}+\Gamma^{\rho}_{\sigma\lambda}\Gamma^{\lambda}_{\chi\eta}-\Gamma^{\rho}_{\chi\lambda}\Gamma^{\lambda}_{\sigma\eta}&\\
&\implies \delta\mathcal{R}^{\rho}_{\eta\sigma\chi}=\partial_{\sigma}\delta\Gamma^{\rho}_{\chi\eta}-\partial_{\chi}\delta\Gamma^{\rho}_{\sigma\eta}+\delta\Gamma^{\rho}_{\sigma\lambda}\Gamma^{\lambda}_{\chi\eta}+\Gamma^{\rho}_{\sigma\lambda}\delta\Gamma^{\lambda}_{\chi\eta}-\delta\Gamma^{\rho}_{\chi\lambda}\Gamma^{\lambda}_{\sigma\eta}-\Gamma^{\rho}_{\chi\lambda}\delta\Gamma^{\lambda}_{\sigma\eta}&\\
&\text{Now, }\nabla_{\sigma}\delta\Gamma^{\rho}_{\chi\eta}=\partial_{\sigma}\delta\Gamma^{\rho}_{\chi\eta}+\Gamma^{\rho}_{\sigma\lambda}\delta\Gamma^{\lambda}_{\chi\eta}-\boldsymbol{\Gamma^{\lambda}_{\sigma\chi}\delta\Gamma^{\rho}_{\lambda\eta}}-\Gamma^{\lambda}_{\sigma\eta}\delta\Gamma^{\rho}_{\chi\lambda}&\\
&\text{and, }\nabla_{\chi}\delta\Gamma^{\rho}_{\sigma\eta}=\partial_{\chi}\delta\Gamma^{\rho}_{\sigma\eta}+\Gamma^{\rho}_{\chi\lambda}\delta\Gamma^{\lambda}_{\sigma\eta}-\boldsymbol{\Gamma^{\lambda}_{\chi\sigma}\delta\Gamma^{\rho}_{\lambda\eta}}-\Gamma^{\lambda}_{\chi\eta}\delta\Gamma^{\rho}_{\sigma\lambda}&\\
&\therefore \delta\mathcal{R}^{\rho}_{\eta\sigma\chi}=\nabla_{\sigma}\delta\Gamma^{\rho}_{\chi\eta}-\nabla_{\chi}\delta\Gamma^{\rho}_{\sigma\eta}&\\
&\delta\mathcal{R}_{\eta\chi}=\delta\mathcal{R}^{\rho}_{\eta\rho\chi}=\nabla_{\rho}\delta\Gamma^{\rho}_{\chi\eta}-\nabla_{\chi}\delta\Gamma^{\rho}_{\rho\eta}&\\
&\text{Now, } \mathcal{R}=g^{\eta\chi}\mathcal{R}_{\eta\chi}\implies \delta\mathcal{R}=\mathcal{R}_{\eta\chi}\delta g^{\eta\chi}+g^{\eta\chi}\delta\mathcal{R}_{\eta\chi}=\mathcal{R}_{\eta\chi}\delta g^{\eta\chi}+g^{\eta\chi}(\nabla_{\rho}\delta\Gamma^{\rho}_{\chi\eta}-\nabla_{\chi}\delta\Gamma^{\rho}_{\rho\eta})&\\
&\implies \delta\mathcal{R}=\mathcal{R}_{\eta\chi}\delta g^{\eta\chi}+g^{\eta\chi}(\nabla_{\rho}\delta\Gamma^{\rho}_{\chi\eta}-\nabla_{\chi}\delta\Gamma^{\rho}_{\rho\eta})&
\end{flalign*}
Since $\delta\Gamma^{\gamma}_{\alpha\beta}$ is the difference of two connections, it should transform as a tensor i.e.
\vspace{-1em}
$$\delta\Gamma^{\gamma}_{\alpha\beta}=\dfrac{1}{2}g^{\gamma\lambda}\left(\nabla_{\alpha}\delta g_{\lambda\beta}+\nabla_{\beta}\delta g_{\lambda\alpha}-\nabla_{\lambda}\delta g_{\alpha\beta}\right)$$
\vspace{-2em}
\begin{flalign*}
&\text{Then }\delta\Gamma^{\rho}_{\chi\eta}=\dfrac{1}{2}g^{\rho\lambda}(\nabla_{\chi}\delta g_{\lambda\eta}+\nabla_{\eta}\delta g_{\lambda\chi}-\nabla_{\lambda}\delta g_{\chi\eta})&\\
&\text{and, } \delta\Gamma^{\rho}_{\rho\eta}=\dfrac{1}{2}g^{\rho\lambda}(\nabla_{\rho}\delta g_{\lambda\eta}+\nabla_{\eta}\delta g_{\lambda\rho}-\nabla_{\lambda}\delta g_{\rho\eta})=\dfrac{1}{2}g^{\rho\lambda}(\nabla_{\eta}\delta g_{\lambda\rho})&
\end{flalign*}
Apply the rule that metric and Kronecker delta (denoted by $\delta$ with an upper and a lower indices) both have no covariant derivative. Use the following identities:
$$\delta g_{\alpha\beta}\delta g^{\alpha\gamma}=\delta^{\gamma}_{\beta} \text{ and }g^{\eta\chi}\delta g_{\eta\chi}=-g_{\eta\chi}\delta g^{\eta\chi}$$
$$\delta g_{\alpha\beta}=-g_{\alpha\eta}g_{\beta\chi}\delta g^{\eta\chi} \text{ and }\delta g^{\alpha\beta}=-g^{\alpha\eta}g^{\beta\chi}\delta g_{\eta\chi}$$
\begin{flalign*}
\therefore \delta\mathcal{R}=&\mathcal{R}_{\eta\chi}\delta g^{\eta\chi}+\dfrac{1}{2}g^{\eta\chi}g^{\rho\lambda}\nabla_{\rho}(\nabla_{\chi}\delta g_{\lambda\eta}+\nabla_{\eta}\delta g_{\lambda\chi}-\nabla_{\lambda}\delta g_{\chi\eta})-\dfrac{1}{2}g^{\eta\chi}g^{\rho\lambda}\nabla_{\chi}(\nabla_{\eta}\delta g_{\lambda\rho})&\\
=&\mathcal{R}_{\eta\chi}\delta g^{\eta\chi}+\dfrac{1}{2}(-g_{\lambda\eta}g^{\rho\lambda}\nabla_{\rho}\nabla_{\chi}\delta g^{\eta\chi}-g_{\lambda\chi}g^{\rho\lambda}\nabla_{\rho}\nabla_{\eta}\delta g^{\eta\chi}+g_{\eta\chi}g^{\rho\lambda}\nabla_{\rho}\nabla_{\lambda}\delta g^{\eta\chi})+\dfrac{1}{2}g^{\eta\chi}g^{\rho\lambda}g_{\lambda\eta}g_{\rho\chi}\nabla_{\chi}\nabla_{\eta}\delta g^{\eta\chi}&\\
=&\mathcal{R}_{\eta\chi}\delta g^{\eta\chi}+\dfrac{1}{2}(-\delta^{\rho}_{\eta}\nabla_{\rho}\nabla_{\chi}\delta g^{\eta\chi}-\delta^{\rho}_{\chi}\nabla_{\rho}\nabla_{\eta}\delta g^{\eta\chi}+g_{\eta\chi}g^{\rho\lambda}\nabla_{\rho}\nabla_{\lambda}\delta g^{\eta\chi})+\dfrac{1}{2}g^{\rho\lambda}g_{\lambda\eta}\delta^{\eta}_{\rho}\nabla_{\chi}\nabla_{\eta}\delta g^{\eta\chi}&\\
=&\mathcal{R}_{\eta\chi}\delta g^{\eta\chi}+\dfrac{1}{2}(-\nabla_{\eta}\nabla_{\chi}\delta g^{\eta\chi}-\nabla_{\eta}\nabla_{\eta}\delta g^{\eta\chi}+g_{\eta\chi}g^{\rho\lambda}\nabla_{\rho}\nabla_{\lambda}\delta g^{\eta\chi})+\dfrac{1}{2}g_{\eta\chi}g^{\rho\lambda}g_{\lambda\eta}\nabla^{\eta}\nabla_{\rho}\delta g^{\eta\chi}&\\
=&\mathcal{R}_{\eta\chi}\delta g^{\eta\chi}+\dfrac{1}{2}(-2\nabla_{\eta}\nabla_{\chi}\delta g^{\eta\chi}+g_{\eta\chi}g^{\rho\lambda}\nabla_{\rho}\nabla_{\lambda}\delta g^{\eta\chi})+\dfrac{1}{2}g_{\eta\chi}g^{\rho\lambda}\nabla_{\lambda}\nabla_{\rho}\delta g^{\eta\chi}&\\
=&\mathcal{R}_{\eta\chi}\delta g^{\eta\chi}-\nabla_{\eta}\nabla_{\chi}\delta g^{\eta\chi}+g_{\eta\chi}g^{\rho\lambda}\nabla_{\rho}\nabla_{\lambda}\delta g^{\eta\chi}&\\
\dfrac{\delta\mathcal{R}}{\delta g^{\eta\chi}}=&\mathcal{R}_{\eta\chi}-\nabla_{\eta}\nabla_{\chi}+g_{\eta\chi}g^{\rho\lambda}\nabla_{\rho}\nabla_{\lambda}&
\end{flalign*}
Thus, $\boldsymbol{\dfrac{\delta\mathcal{R}}{\delta g^{\mu\nu}}=\mathcal{R}_{\mu\nu}-\nabla_{\mu}\nabla_{\nu}+g_{\mu\nu}\Box}$ where $\Box=g^{\rho\lambda}\nabla_{\rho}\nabla_{\lambda}$\\
Hence, the field equation in $f(\mathcal{R})$ gravity is: $\boldsymbol{(\mathcal{R}_{\mu\nu}-\nabla_{\mu}\nabla_{\nu}+g_{\mu\nu}\Box\,)f_{\mathcal{R}}-\dfrac{1}{2}f(\mathcal{R})g_{\mu\nu}=\kappa T_{\mu\nu}}$

\subsection*{Variation of Einstein tensor}
Einstein tensor (i.e. trace-reversed Ricci tensor), $\mathcal{G}_{\mu\nu}=\mathcal{R}_{\mu\nu}-\dfrac{1}{2}\mathcal{R}g_{\mu\nu}$\\
$\implies\delta \mathcal{G}_{\mu\nu}=\delta \mathcal{R}_{\mu\nu}-\dfrac{1}{2}g_{\mu\nu}\delta\mathcal{R}-\dfrac{1}{2}\mathcal{R}\delta g_{\mu\nu}$\\
$\implies\delta \mathcal{G}_{\mu\nu}=(\nabla_{\sigma}\delta\Gamma^{\sigma}_{\nu\mu}-\nabla_{\nu}\delta\Gamma^{\sigma}_{\sigma\mu})-\dfrac{1}{2}g_{\mu\nu}\left[\mathcal{R}_{\alpha\beta}\delta g^{\alpha\beta}+g^{\alpha\beta}\delta \mathcal{R}_{\alpha\beta}\right]-\dfrac{1}{2}\mathcal{R}\delta g_{\mu\nu}$\\
$\implies\delta \mathcal{G}_{\mu\nu}=\dfrac{1}{4}g^{\mu\nu}\nabla_{\mu}\nabla_{\nu}\delta g_{\mu\nu}-\Box\,\delta g_{\mu\nu}-\dfrac{1}{2}g_{\mu\nu}\left[\mathcal{R}_{\alpha\beta}\delta g^{\alpha\beta}-\nabla_{\alpha}\nabla_{\beta}\delta g^{\alpha\beta}+g_{\alpha\beta}\Box\,\delta g^{\alpha\beta}\right]-\dfrac{1}{2}\mathcal{R}\delta g_{\mu\nu}$\\\\
Since, $g^{\mu\nu}(\nabla_{\rho}\delta\Gamma^{\rho}_{\nu\mu}-\nabla_{\nu}\delta\Gamma^{\rho}_{\rho\mu})=-\nabla_{\mu}\nabla_{\nu}\delta g^{\mu\nu}+g_{\mu\nu}g^{\rho\lambda}\nabla_{\rho}\nabla_{\lambda}\delta g^{\mu\nu}$\\
$\implies \nabla_{\rho}\delta\Gamma^{\rho}_{\nu\mu}-\nabla_{\nu}\delta\Gamma^{\rho}_{\rho\mu}=-\dfrac{1}{4}g_{\mu\nu}\nabla_{\mu}\nabla_{\nu}\delta g^{\mu\nu}+\dfrac{1}{4}g_{\mu\nu}g_{\mu\nu}g^{\rho\lambda}\nabla_{\rho}\nabla_{\lambda}\delta g^{\mu\nu}$\\
$\implies \delta \mathcal{R}_{\mu\nu}=\dfrac{1}{4}g^{\mu\nu}\nabla_{\mu}\nabla_{\nu}\delta g_{\mu\nu}-\Box\,\delta g_{\mu\nu}$

\subsection*{Trace-reversed Field Equations}
$\mathcal{R}_{\mu\nu}-\dfrac{1}{2}\mathcal{R}g_{\mu\nu}=\kappa T_{\mu\nu}$\\\\
Take the trace of the above equation i.e. multiply throughout by $g^{\mu\nu}$\\\\
$\implies \mathcal{R}-2\mathcal{R}=\kappa T\implies \mathcal{R}=-\kappa T\qquad [\,\because g^{\mu\nu}\mathcal{R}_{\mu\nu}=\mathcal{R} \text{ and }g^{\mu\nu}g_{\mu\nu}=4\,]$\\\\
Put this into the field equations.\\\\
$\mathcal{R}_{\mu\nu}-\dfrac{1}{2}(-\kappa T)g_{\mu\nu}=\kappa T_{\mu\nu}\implies \mathcal{R}_{\mu\nu}=\kappa\Big(T_{\mu\nu}-\dfrac{1}{2}T\Big)$\\\\
which is the desired trace-reversed form of the field equations.

\subsection*{Field equations (FE) from energy-momentum conservation}
Ansatz of FE: $\quad\mathcal{R}_{\mu\nu}=\kappa T_{\mu\nu}$\\\\
However, this is not consistent with energy-momentum conservation (i.e. $\nabla_{\mu}T^{\mu\nu}=0$) as $\nabla_{\mu}\mathcal{R}_{\mu\nu}\neq 0$.\\This is a similar situation as in the derivation of Maxwell's fourth equation of electromagnetism.\\\\
Second (differential) Bianchi identity: $\nabla_{\sigma}\mathcal{R}_{\mu\nu\lambda\rho}+\nabla_{\rho}\mathcal{R}_{\mu\nu\sigma\lambda}+\nabla_{\lambda}\mathcal{R}_{\mu\nu\rho\sigma}=0$\\\\
$\implies \nabla_{\sigma}\mathcal{R}^{\lambda}_{\;\nu\lambda\rho}+\nabla_{\rho}\mathcal{R}^{\lambda}_{\;\nu\sigma\lambda}+\nabla_{\lambda}\mathcal{R}^{\lambda}_{\;\nu\rho\sigma}=0\quad[\,\text{contracting with }g^{\lambda\mu}\,]$\\ $\implies \nabla_{\sigma}\mathcal{R}^{\lambda}_{\;\nu\lambda\rho}-\nabla_{\rho}\mathcal{R}^{\lambda}_{\;\nu\lambda\sigma}+\nabla_{\lambda}\mathcal{R}^{\lambda}_{\;\nu\rho\sigma}=0\quad[\text{using antisymmetry property of the Riemann tensor}]$\\
$\implies \nabla_{\sigma}\mathcal{R}_{\nu\rho}-\nabla_{\rho}\mathcal{R}_{\nu\sigma}+\nabla_{\lambda}\mathcal{R}^{\lambda}_{\;\nu\rho\sigma}=0\quad[\text{using definition of the Ricci tensor}]$\\
$\implies \nabla_{\sigma}\mathcal{R}^{\rho}_{\;\rho}-\nabla_{\rho}\mathcal{R}^{\rho}_{\;\sigma}-\nabla_{\lambda}\mathcal{R}^{\lambda\rho}_{\;\;\rho\sigma}=0\quad[\text{contracting with }g^{\nu\rho}\text{ and using antisymmetry property of Riemann tensor}]$\\
$\implies \nabla_{\sigma}\mathcal{R}-\nabla_{\rho}\mathcal{R}^{\rho}_{\sigma}-\nabla_{\lambda}\mathcal{R}^{\lambda}_{\;\sigma}=0\quad[\text{using definition of the Ricci tensor and the Ricci scalar}]$\\
$\implies \nabla_{\sigma}\mathcal{R}-\nabla_{\lambda}\mathcal{R}^{\lambda}_{\sigma}-\nabla_{\lambda}\mathcal{R}^{\lambda}_{\;\sigma}=0\quad[\text{relabelling dummy index}]$\\
$\implies \nabla_{\lambda}\delta^{\lambda}_{\;\sigma}\mathcal{R}-2\,\nabla_{\lambda}\mathcal{R}^{\lambda}_{\sigma}=0\quad[\text{changing index}]$\\
$\implies \nabla_{\lambda}g^{\lambda\nu}\mathcal{R}-2\,\nabla_{\lambda}\mathcal{R}^{\lambda\nu}=0\quad[\text{contracting with }g^{\nu\sigma}]$\\
$\implies \nabla_{\lambda}(\mathcal{R}^{\lambda\nu}-\dfrac{1}{2}g^{\lambda\nu}\mathcal{R})=0\quad[\text{rearranging terms}]$\\
$\implies \nabla_{\mu}(\mathcal{R}^{\mu\nu}-\dfrac{1}{2}g^{\mu\nu}\mathcal{R})=0\quad[\text{relabelling dummy index}]$\\\\
$\implies \nabla_{\mu}(\mathcal{R}^{\mu\nu}-\dfrac{1}{2}g^{\mu\nu}\mathcal{R})=\kappa T_{\mu\nu}\quad[\because \nabla_{\mu}T^{\mu\nu}=0]$\\\\
$\implies\mathcal{R}^{\mu\nu}-\dfrac{1}{2}g^{\mu\nu}\mathcal{R}=\kappa T^{\mu\nu}\Leftrightarrow\boldsymbol{\mathcal{R}_{\mu\nu}-\dfrac{1}{2}g_{\mu\nu}\mathcal{R}=\kappa T_{\mu\nu}}$

\subsection*{GR in Newtonian limit: Evaluating $\kappa$}
Field Equations (trace-reversed): $\mathcal{R}_{\mu\nu}=\kappa\Big(T_{\mu\nu}-\dfrac{1}{2}g_{\mu\nu}T\Big)$\\
Geodesic equation: $\dfrac{d^2 x^{\mu}}{d\tau^2}=-\Gamma^{\mu}_{\rho\sigma}\dfrac{d x^{\rho}}{d\tau}\dfrac{d x^{\sigma}}{d\tau}$\\\\
Correspondingly, in Newtonian gravitation, the equations are:
\begin{flalign*}
&\text{Poisson's equation for gravity: }\nabla^2\Phi=4\pi G\rho&\\
&\text{Newton's law of motion in gravity: }\dfrac{d^2 x^i}{dt^2}=-\nabla\Phi&
\end{flalign*}
In Newtonian limit, the assumptions are\\\\
Weak field: $g_{\mu\nu}=\eta_{\mu\nu}+h_{\mu\nu}\implies g^{\mu\nu}\approx\eta^{\mu\nu}-\eta^{\mu\alpha}\eta^{\nu\beta}h_{\alpha\beta}$\\\\
Remark: The inverse metric is such so that $g^{\mu\lambda}g_{\lambda\nu}\approx \delta^{\mu}_{\;\nu}$ as shown below.\\\\
$\big[\,g^{\mu\lambda}g_{\lambda\nu}=(\eta^{\mu\lambda}-\eta^{\mu\alpha}\eta^{\lambda\beta}h_{\alpha\beta})(\eta_{\lambda\nu}+h_{\lambda\nu})\approx \delta^{\mu}_{\;\nu}+\eta^{\mu\lambda}h_{\lambda\nu}-\eta^{\mu\alpha}\eta^{\lambda\beta}h_{\alpha\beta}\eta_{\lambda\nu}=\delta^{\mu}_{\;\nu}+\eta^{\mu\lambda}h_{\lambda\nu}-\delta^{\beta}_{\;\nu}\eta^{\mu\alpha}h_{\alpha\beta}\\\\
=\delta^{\mu}_{\;\nu}+\eta^{\mu\lambda}h_{\lambda\nu}-\eta^{\mu\alpha}h_{\alpha\nu}=\delta^{\mu}_{\;\nu}+\eta^{\mu\lambda}h_{\lambda\nu}-\eta^{\mu\lambda}h_{\lambda\nu}=\delta^{\mu}_{\;\nu}\,\big]$\\\\
Static field: $\partial_0 g_{\mu\nu}=0\implies \partial^0 g_{\mu\nu}=0$\\
Slow motion: $\dfrac{dx^i}{d\tau}\ll c\implies \dfrac{dx^i}{d\tau}\ll\dfrac{dt}{d\tau}\quad\text{so that}\quad\dfrac{d^2 x^{\mu}}{d\tau^2}\approx -\Gamma^{\mu}_{00}\bigg(\dfrac{d x^0}{d\tau}\bigg)^2=-\Gamma^{\mu}_{00}\bigg(\dfrac{dt}{d\tau}\bigg)^2$\\
Now, $\Gamma^{\mu}_{00}=\dfrac{1}{2}g^{\mu\lambda}(\partial_0 g_{\lambda 0}+\partial_0 g_{0\lambda}-\partial_{\lambda} g_{00})=-\dfrac{1}{2}g^{\mu\lambda}\partial_{\lambda} g_{00}\quad[\because \partial_0 g_{\mu\nu}=0]\quad=-\dfrac{1}{2}\partial^{\mu} h_{00}\quad[\because \partial_{\lambda}\eta_{00}=0]$\\
$\therefore \dfrac{d^2 x^{\mu}}{d\tau^2}\approx \dfrac{1}{2}\bigg(\dfrac{dt}{d\tau}\bigg)^2\partial^{\mu} h_{00}$\\\\
The time component of the above equation (i.e. $\mu=0$):\\\\
$\dfrac{d^2 t}{d\tau^2}\approx \dfrac{1}{2}\bigg(\dfrac{dt}{d\tau}\bigg)^2\partial^0 h_{00}=0\quad[\because \partial^0 h_{00}=0\,]\quad\implies \tau=t$\\\\
The space component of the above equation (i.e. $\mu=i$):\\\\
$\dfrac{d^2 x^i}{d\tau^2}\approx \dfrac{1}{2}\bigg(\dfrac{dt}{d\tau}\bigg)^2\partial^i h_{00}=\dfrac{1}{2}\partial^i h_{00}=\dfrac{1}{2}\partial_i h_{00}\implies \boldsymbol{\dfrac{d^2 x^i}{dt^2}=-\partial_i \Big(-\dfrac{1}{2}h_{00}\Big)=-\partial_i \Phi}\quad\Big[\text{where }\Phi=-\dfrac{1}{2}h_{00}\Big]$\\
$[\,\because \partial^i h_{00}=g^{ij}\partial_j h_{00}=(\eta^{ij}-\eta^{ia}\eta^{jb}h_{ab})\partial_j h_{00}\approx \eta^{ij}\partial_j h_{00}=\eta^{ii}\partial_i h_{00}=\partial_i h_{00}\,]$\\\\
Now, $T_{\mu\nu}=\text{diag}(\rho c^4,0,0,0)\quad\text{so that}\quad T=g^{\mu\nu}T_{\mu\nu}=(\eta^{\mu\nu}-\eta^{\mu\alpha}\eta^{\nu\beta}h_{\alpha\beta})T_{\mu\nu}=(\eta^{00}-\eta^{0\alpha}\eta^{0\beta}h_{\alpha\beta})T_{00}$\\\\
$\implies T=(\eta^{00}-\eta^{00}\eta^{00}h_{00})T_{00}=(-c^{-2}-c^{-4} h_{00})\rho c^4=(-c^2-h_{00})\rho$\\\\
Next, $\mathcal{R}_{\mu\nu}=\kappa\Big(T_{\mu\nu}-\dfrac{1}{2}g_{\mu\nu}T\Big)\implies \mathcal{R}_{00}=\kappa\Big(T_{00}-\dfrac{1}{2}g_{00}T\Big)=\kappa\Big(T_{00}-\dfrac{1}{2}(\eta_{00}+h_{00})T\Big)$\\
$=\kappa\Big(\rho c^4-\dfrac{1}{2}(-c^2+h_{00})(-c^2-h_{00})\rho\Big)\quad[\because g_{00}=\eta_{00}+h_{00}]\quad \implies \mathcal{R}_{00}=\dfrac{1}{2}\kappa \rho c^4$\\
$\implies \dfrac{1}{2}\kappa \rho c^4=R^{\lambda}_{0\lambda 0}=\partial_{\lambda}\Gamma^{\lambda}_{00}-\partial_0\Gamma^{\lambda}_{\lambda 0}+\Gamma^{\lambda}_{\lambda\sigma}\Gamma^{\sigma}_{00}-\Gamma^{\lambda}_{0\sigma}\Gamma^{\sigma}_{\lambda 0}\approx \partial_{\lambda}\Gamma^{\lambda}_{00}$\\\\
$[\,\because \text{other terms are zero or negligible by static and weak field assumptions}\,]$\\
$\implies \dfrac{1}{2}\kappa \rho c^4=\partial_0\Gamma^0_{00}+\partial_i\Gamma^i_{00}=\partial_i\Gamma^i_{00}=\dfrac{1}{2}\partial_i g^{i\lambda}(\partial_0 g_{\lambda 0}+\partial_0 g_{0\lambda}-\partial_{\lambda} g_{00})=-\dfrac{1}{2}\partial_i g^{i\lambda}\partial_{\lambda} g_{00}=-\dfrac{1}{2}\partial_i \partial^i h_{00}$\\
$\implies \dfrac{1}{2}\kappa \rho c^4=\partial_i \partial^i \Big(-\dfrac{1}{2}h_{00}\Big)=\partial_i \partial^i \Phi=4\pi G\rho\implies \boldsymbol{\kappa=\dfrac{8\pi G}{c^4}}\approx 2\times 10^{-43}\text{\,m}^{-1} \text{\,kg}^{-1}\text{\,s}^2$

\subsection*{Minimizing the Residual (analytically)}
\vspace{-1em}
\begin{flalign*}
\text{res}^2=&\dfrac{1}{N}\sum_{i}\sum_{j}(\log{p_j}-\log{K_i}+\Gamma_i\log{\rho_j})^2&\\
=&\dfrac{1}{N}\bigg[\sum_{j_{1}}(\log{p_{j_{1}}}-\log{K_1}+\Gamma_1\log{\rho_{j_{1}}})^2+\sum_{j_{2}}(\log{p_{j_{2}}}-\log{K_2}+\Gamma_2\log{\rho_{j_{2}}})^2&\\
&+\sum_{j_{3}}(\log{p_{j_{3}}}-\log{K_3}+\Gamma_3\log{\rho_{j_{3}}})^2\bigg]&\\
=&\dfrac{1}{N}\bigg[\sum_{j_{1}}(\log{p_{j_{1}}}-(\log{p_1}-\Gamma_1 \log{\rho_1})+\Gamma_1\log{\rho_{j_{1}}})^2+\sum_{j_{2}}(\log{p_{j_{2}}}-(\log{p_1}-\Gamma_2 \log{\rho_1})+\Gamma_2\log{\rho_{j_{2}}})^2&\\
&+\sum_{j_{3}}(\log{p_{j_{3}}}-(\log{p_1}+(\log{\rho_2}-\log{\rho_1})\Gamma_2-\Gamma_3\log{\rho_2})+\Gamma_3\log{\rho_{j_{3}}})^2\bigg]&
\end{flalign*}
where $i$ labels the piecewise polytrope segments and $j$ labels the data points falling in the density range spanned by the $i^\text{th}$ segment.\\
To minimize the residual, we need $\dfrac{\partial (\text{res}^2)}{\partial \Gamma_1}=0$, $\dfrac{\partial (\text{res}^2)}{\partial \Gamma_2}=0$, $\dfrac{\partial (\text{res}^2)}{\partial \Gamma_3}=0$ and $\dfrac{\partial (\text{res}^2)}{\partial (\log{p_1})}=0$
\begin{flalign*}
\text{Now, }&\dfrac{\partial (\text{res}^2)}{\partial \Gamma_1}=0&\\
\implies&\sum_{j_{1}}(\log{p_{j_{1}}}-\log{p_1}+\Gamma_1 \log{\rho_1}+\Gamma_1\log{\rho_{j_{1}}})(\log{\rho_1}+\log{\rho_{j_{1}}})=0&\\
\implies&\sum_{j_{1}}(\log{p_{j_{1}}}-\log{p_1})(\log{\rho_1}+\log{\rho_{j_{1}}})+\Gamma_1\sum_{j_{1}}(\log{\rho_1}+\log{\rho_{j_{1}}})^2=0&\\
\implies&\Gamma_1=\dfrac{-\sum_{j_{1}}(\log{p_{j_{1}}}-\log{p_1})(\log{\rho_1}+\log{\rho_{j_{1}}})}{\sum_{j_{1}}(\log{\rho_1}+\log{\rho_{j_{1}}})^2}&
\end{flalign*}
\begin{flalign*}
\text{And, }&\dfrac{\partial (\text{res}^2)}{\partial \Gamma_2}=0&\\
\implies&\sum_{j_{2}}(\log{p_{j_{2}}}-\log{p_1}+\Gamma_2 \log{\rho_1}+\Gamma_2\log{\rho_{j_{2}}})(\log{\rho_1}+\log{\rho_{j_{2}}})&\\
&-\sum_{j_{3}}(\log{p_{j_{3}}}-\log{p_1}-(\log{\rho_2}-\log{\rho_1})\Gamma_2+\Gamma_3\log{\rho_2}+\Gamma_3\log{\rho_{j_{3}}})(\log{\rho_2}-\log{\rho_1})=0&\\
\implies&\sum_{j_{2}}(\log{p_{j_{2}}}-\log{p_1})(\log{\rho_1}+\log{\rho_{j_{2}}})+\Gamma_2\sum_{j_{2}}(\log{\rho_1}+\log{\rho_{j_{2}}})^2-(\log{\rho_2}-\log{\rho_1})\sum_{j_{3}}(\log{p_{j_{3}}}-\log{p_1})&\\
&+\Gamma_2(\log{\rho_2}-\log{\rho_1})^2n(j_3)+\Gamma_3(\log{\rho_2}-\log{\rho_1})\sum_{j_{3}}(\log{\rho_2}+\log{\rho_{j_{3}}})=0&\\
\implies&\Gamma_2\bigg[\sum_{j_{2}}(\log{\rho_1}+\log{\rho_{j_{2}}})^2+(\log{\rho_2}-\log{\rho_1})^2n(j_3)\bigg]
+\Gamma_3(\log{\rho_2}-\log{\rho_1})\sum_{j_{3}}(\log{\rho_2}+\log{\rho_{j_{3}}})&\\
&=-\sum_{j_{2}}(\log{p_{j_{2}}}-\log{p_1})(\log{\rho_1}+\log{\rho_{j_{2}}})+(\log{\rho_2}-\log{\rho_1})\sum_{j_{3}}(\log{p_{j_{3}}}-\log{p_1})&
\end{flalign*}
\begin{flalign*}
\text{With, }&\dfrac{\partial (\text{res}^2)}{\partial \Gamma_3}=0&\\
\implies&\sum_{j_{3}}(\log{p_{j_{3}}}-\log{p_1}-(\log{\rho_2}-\log{\rho_1})\Gamma_2+\Gamma_3\log{\rho_2}+\Gamma_3\log{\rho_{j_{3}}})(\log{\rho_2}+\log{\rho_{j_{3}}})=0&\\
\implies&\sum_{j_{3}}(\log{p_{j_{3}}}-\log{p_1})(\log{\rho_2}+\log{\rho_{j_{3}}})-\Gamma_2(\log{\rho_2}-\log{\rho_1})\sum_{j_{3}}(\log{\rho_2}+\log{\rho_{j_{3}}})&\\
&+\Gamma_3\sum_{j_{3}}(\log{\rho_2}+\log{\rho_{j_{3}}})^2=0&\\
\implies&\Gamma_2(\log{\rho_2}-\log{\rho_1})\sum_{j_{3}}(\log{\rho_2}+\log{\rho_{j_{3}}})-\Gamma_3\sum_{j_{3}}(\log{\rho_2}+\log{\rho_{j_{3}}})^2&\\
&=\sum_{j_{3}}(\log{p_{j_{3}}}-\log{p_1})(\log{\rho_2}+\log{\rho_{j_{3}}})&
\end{flalign*}
\begin{flalign*}
\therefore\,\,&\Gamma_2\bigg[\sum_{j_{2}}(\log{\rho_1}+\log{\rho_{j_{2}}})^2\sum_{j_{3}}(\log{\rho_2}+\log{\rho_{j_{3}}})^2+(\log{\rho_2}-\log{\rho_1})^2n(j_3)\sum_{j_{3}}(\log{\rho_2}+\log{\rho_{j_{3}}})^2&\\
&+(\log{\rho_2}-\log{\rho_1})^2\Big(\sum_{j_{3}}(\log{\rho_2}+\log{\rho_{j_{3}}})\Big)^2\bigg]&\\
&=-\sum_{j_{2}}(\log{p_{j_{2}}}-\log{p_1})(\log{\rho_1}+\log{\rho_{j_{2}}})\sum_{j_{3}}(\log{\rho_2}+\log{\rho_{j_{3}}})^2&\\
&+(\log{\rho_2}-\log{\rho_1})\sum_{j_{3}}(\log{p_{j_{3}}}-\log{p_1})\sum_{j_{3}}(\log{\rho_2}+\log{\rho_{j_{3}}})^2&\\
&+(\log{\rho_2}-\log{\rho_1})\sum_{j_{3}}(\log{p_{j_{3}}}-\log{p_1})(\log{\rho_2}+\log{\rho_{j_{3}}})\sum_{j_{3}}(\log{\rho_2}+\log{\rho_{j_{3}}})&\\
&\implies \Gamma_2=\dfrac{\splitdfrac{\splitdfrac{-\sum_{j_{2}}(\log{p_{j_{2}}}-\log{p_1})(\log{\rho_1}+\log{\rho_{j_{2}}})\sum_{j_{3}}(\log{\rho_2}+\log{\rho_{j_{3}}})^2}{+(\log{\rho_2}-\log{\rho_1})\sum_{j_{3}}(\log{p_{j_{3}}}-\log{p_1})\sum_{j_{3}}(\log{\rho_2}+\log{\rho_{j_{3}}})^2}}{+(\log{\rho_2}-\log{\rho_1})\sum_{j_{3}}(\log{p_{j_{3}}}-\log{p_1})(\log{\rho_2}+\log{\rho_{j_{3}}})\sum_{j_{3}}(\log{\rho_2}+\log{\rho_{j_{3}}})}}{\splitdfrac{\sum_{j_{2}}(\log{\rho_1}+\log{\rho_{j_{2}}})^2\sum_{j_{3}}(\log{\rho_2}+\log{\rho_{j_{3}}})^2+(\log{\rho_2}-\log{\rho_1})^2n(j_3)\sum_{j_{3}}(\log{\rho_2}+\log{\rho_{j_{3}}})^2}{+(\log{\rho_2}-\log{\rho_1})^2\big(\sum_{j_{3}}(\log{\rho_2}+\log{\rho_{j_{3}}})\big)^2}}&
\end{flalign*}
\begin{flalign*}
\text{And, }&\Gamma_3(\log{\rho_2}-\log{\rho_1})^2\bigg(\sum_{j_{3}}(\log{\rho_2}+\log{\rho_{j_{3}}})\bigg)^2&\\
&+\Gamma_3\sum_{j_{3}}(\log{\rho_2}+\log{\rho_{j_{3}}})^2\bigg[\sum_{j_{2}}(\log{\rho_1}+\log{\rho_{j_{2}}})^2+(\log{\rho_2}-\log{\rho_1})^2n(j_3)\bigg]&\\
&=-\sum_{j_{3}}(\log{p_{j_{3}}}-\log{p_1})(\log{\rho_2}+\log{\rho_{j_{3}}})\bigg[\sum_{j_{2}}(\log{\rho_1}+\log{\rho_{j_{2}}})^2+(\log{\rho_2}-\log{\rho_1})^2n(j_3)\bigg]&\\
&-(\log{\rho_2}-\log{\rho_1})\sum_{j_{2}}(\log{p_{j_{2}}}-\log{p_1})(\log{\rho_1}+\log{\rho_{j_{2}}})\sum_{j_{3}}(\log{\rho_2}+\log{\rho_{j_{3}}})&\\
&+(\log{\rho_2}-\log{\rho_1})^2\sum_{j_{3}}(\log{p_{j_{3}}}-\log{p_1})\sum_{j_{3}}(\log{\rho_2}+\log{\rho_{j_{3}}})&\\
&\implies \Gamma_3=\dfrac{\splitdfrac{\splitdfrac{-\sum_{j_{3}}(\log{p_{j_{3}}}-\log{p_1})(\log{\rho_2}+\log{\rho_{j_{3}}})\big[\sum_{j_{2}}(\log{\rho_1}+\log{\rho_{j_{2}}})^2+(\log{\rho_2}-\log{\rho_1})^2n(j_3)\big]}{-(\log{\rho_2}-\log{\rho_1})\sum_{j_{2}}(\log{p_{j_{2}}}-\log{p_1})(\log{\rho_1}+\log{\rho_{j_{2}}})\sum_{j_{3}}(\log{\rho_2}+\log{\rho_{j_{3}}})}}{+(\log{\rho_2}-\log{\rho_1})^2\sum_{j_{3}}(\log{p_{j_{3}}}-\log{p_1})\sum_{j_{3}}(\log{\rho_2}+\log{\rho_{j_{3}}})}}{\splitdfrac{(\log{\rho_2}-\log{\rho_1})^2\big(\sum_{j_{3}}(\log{\rho_2}+\log{\rho_{j_{3}}})\big)^2}{+\sum_{j_{3}}(\log{\rho_2}+\log{\rho_{j_{3}}})^2\big[\sum_{j_{2}}(\log{\rho_1}+\log{\rho_{j_{2}}})^2+(\log{\rho_2}-\log{\rho_1})^2n(j_3)\big]}}
\end{flalign*}
\begin{flalign*}
\text{Now, }&\dfrac{\partial (\text{res}^2)}{\partial (\log{p_1})}=0&\\
\implies&\sum_{j_{1}}(\log{p_{j_{1}}}-(\log{p_1}-\Gamma_1 \log{\rho_1})+\Gamma_1\log{\rho_{j_{1}}})+\sum_{j_{2}}(\log{p_{j_{2}}}-(\log{p_1}-\Gamma_2 \log{\rho_1})+\Gamma_2\log{\rho_{j_{2}}})&\\
&+\sum_{j_{3}}(\log{p_{j_{3}}}-(\log{p_1}+(\log{\rho_2}-\log{\rho_1})\Gamma_2-\Gamma_3\log{\rho_2})+\Gamma_3\log{\rho_{j_{3}}})=0&
\end{flalign*}
where $\Gamma_1$, $\Gamma_2$ and $\Gamma_3$ are functions of $\log{p_1}$. Solving the above equation, we get $\log{p_1}$ and the adiabatic indices for three pieces for which the residual is minimum.


\end{document}